\documentclass{aastex63}

\shorttitle{Experiments on the electrostatic transport of charged anorthite particles et al.}
\shortauthors{Gan et al.}

\begin{document}
%\linenumbers
\title{Experiments on the Electrostatic Transport of Charged Anorthite Particles under Electron Beam Irradiation}

\correspondingauthor{Xiaoping Zhang}
\email{ganhong06@gmail.com, xpzhangnju@gmail.com}

\author[0000-0001-7667-0498]{Hong Gan}
\affiliation{State Key Laboratory of Lunar and Planetary Sciences, Macau University of Science and Technology, Macau, PR China}
\affiliation{CNSA Macau Center for Space Exploration and Science, Macau, PR China}
\affiliation{Analyzing and Testing Center, Guizhou Institute of Technology, Guiyang, China}

\author{Xiaoping Zhang}
\affiliation{State Key Laboratory of Lunar and Planetary Sciences, Macau University of Science and Technology, Macau, PR China}
\affiliation{CNSA Macau Center for Space Exploration and Science, Macau, PR China}

\author{Xiongyao Li}
\affiliation{Center for Lunar and Planetary Sciences, Institute of Geochemistry, Chinese Academy of Sciences, Guiyang, China}
\affiliation{Center for Excellence in Comparative Planetology, Chinese Academy of Sciences, Hefei, China}

\author{Hong Jin}
\affiliation{Center for Lunar and Planetary Sciences, Institute of Geochemistry, Chinese Academy of Sciences, Guiyang, China}

\author{Lianghai Xie}
\affiliation{National Space Science Center, Chinese Academy of Sciences, Beijing, China}

\author{Yongliao Zou}
\affiliation{National Space Science Center, Chinese Academy of Sciences, Beijing, China}

\begin{abstract}

To reveal the effect of secondary electron emission on the charging properties of a surface covered by micron-sized insulating dust particles and the migration characteristics of these particles, for the first time, we used a laser Doppler method to measure the diameters and velocities of micron-sized anorthite particles under electron beam irradiation with an incident energy of 350 eV. Here, anorthite particles are being treated as a proxy for lunar regolith. We experimentally confirm that the vertical transport of anorthite particles is always dominant, although the horizontal transport occurs. In our experiments, some anorthite particles were observed to have large vertical velocities up to 9.74 m~s$^{-1}$ at the measurement point. The upper boundary of the vertical velocities $V_{\rm{z}}$ of these high-speed anorthite particles are well constrained by its diameter $D$, that is, $V_{\rm{z}}^2$ linearly depends on $D^{-2}$. These velocity-diameter data provide strong constraints on the dust charging and transportation mechanisms. The shared charge model could not explain the observed velocity-diameter data. Both the isolated charge model and patched charge model appear to require a large dust charging potential of $-$350 to $-$78 V to reproduce the observed data. The micro-structures of the dusty surface may play an important role in producing this charging potential and in understanding the pulse migration phenomenon observed in our experiment. The presented results and analysis in this paper are helpful for understanding the dust charging and electrostatic transport mechanisms in airless celestial bodies such as the Moon and asteroids in various plasma conditions. 

\end{abstract}

\keywords{Moon, lunar dust --- anorthite --- electrostatic transport --- secondary electron emission}

\section{Introduction} \label{sec:intro}

Dust, one of the common components in space, always interacts with ambient plasma and various radiations and can then be charged and transported. Multiple observations of the lunar horizon glow (LHG) by Surveyor 5, 6, and 7 provide the possible evidence of the electrostatic dust transport phenomenon on the lunar surface, which can be interpreted as forward scattering light from a cloud of micron-sized dust grains at heights within 1 m above the surface \citep{criswell1973horizon, rennilson1974surveyor, colwell2007lunar}. Subsequently, high altitude LHG observed by astronauts from orbit above the Moon was also reported during Apollo missions, which was consistent with a population of submicron-sized dust in the lunar exosphere with a scale height of approximately 10 km \citep{mccoy1976photometric, zook1991large}. Although, initially proposed to be caused by electrostatic transport, later analysis revealed that high altitude LHG events were far less frequent than indicated by earlier reports and probably produced by some other mechanism \citep{glenar2011reanalysis}. In addition, the possible evidence for electrostatic dust transport was discovered on airless bodies, such as ``dust ponds'' in craters on Asteroid 433 Eros \citep{robinson2001nature, colwell2005Dust}. To measure the dust flux variations, the Lunar Ejecta and Meteorites (LEAM) experiment was deployed on the Moon during the Apollo 17 mission, and lunar fines movement was detected earlier \citep{berg1976lunar}. However, the results were highly controversial\citep{o2011review, laursen2011apollo}. Although the Lunar Atmosphere and Dust Environment Explorer (LADEE) mission orbited the Moon from October 2013 to April 2014, found no electrostatically lifted dust grains in the altitude range of 3-250 km above the lunar surface \citep{szalay2015search, horanyi2015permanent}, the latest finding of five dust enhancement events during the LADEE mission again provide evidence of electrostatic dust lifting from the lunar surface \citep{xie2020lunar}. Moreover, according to distinct reflectance features of lunar rocks and regolith observed by the Chang'E-3 mission launched on 2 December 2013, long-term dust activity below approximately 28 cm above the lunar surface has been identified \citep{yan2019weak}. These results provides strong evidence for the electrostatic migration of lunar dust in the vicinity of the lunar surface.

Dust grains and the lunar surface are charged mainly due to solar UV radiation and solar wind plasma \citep{horanyi1996charged, stubbs2014dependence}. The stable surface potential on a dust grain or the lunar surface could be determined according to the balance between various currents, such as the currents from solar wind electrons, solar wind protons, secondary electrons, and photoelectrons. In particular, the interaction between electrons and dust grains is complex but interesting. Dust particles can charge negatively due to electron attachment or positively as a result of secondary electron emission (SEE). This means that, in a given environment, two identical grains may have different amounts of charges or even opposite charges \citep{vaverka2013numerical}, or one particle can have opposite charges on different parts of its surface because of its rather low conductivity. Thus, dust grains can be charged to multivalued potentials corresponding to different charging conditions and/or histories \citep{meyer1982flip}. For example, the lunar surface potential changes when the Moon travels through the terrestrial bow shock. The lunar surface potential obtained in the upstream foreshock region is different from that obtained in the downstream magnetosheath region due to different environmental conditions \citep{collier2011lunar}. For electrical insulators, such as most lunar surface substances, no accurate theory can be used in astronomical applications due to the mobility and unknown lifetimes of existing charges \citep{dukes2013secondary}.

As one of the main charging mechanisms of lunar surface, the interaction between electrons and dust grains is closely related to the surface potential distribution and the dust transport characteristics. This interaction plays a prominent role on the whole surface of the Moon. A series of experiments have demonstrated the transport of dust particles exposed to electron beams, but the energies of incident electrons were below 120 eV \citep{wang2010investigation, wang2011dust, wang2016dust}. Dust particles cannot be liberated by the plasma sheath electric field alone but can lift off the surface when exposed to an electron beam with or without a background plasma \citep{wang2016dust}, which indicates the importance of electrons in dust motion. Based on these laboratory results, a new ``patched charge model'' has been proposed to explain that in typical plasma conditions, the net electrostatic force according to the shared model is not enough to overcome gravity \citep{wang2016dust}. Then, when exposed to a 120-eV electron beam, the lifting rate of dust grains over time \citep{hood2018laboratory} and the initial launch velocities of these dust grains \citep{carroll2020laboratory} have been discussed. The results show that such intense electrostatic dust lifting can only last over a short duration and that the launch velocity of electrostatically lifted dust is related to its particle size and shape. A recent laboratory investigation on silica dust lifting has been performed under an electron beam with 450 eV energy to identify the importance of the various properties, such as the packing density of the dust samples and the horizontal electric field \citep{orger2019experimental}, and the launch angle distribution of lifted dust particles \citep{orger2021experimental}.

Besides these laboratory studies, the in situ Charged Particle Lunar Environment Experiment (CPLEE) during the Apollo 14 mission has measured the fluxes of charged particles with energies ranging from 50 eV to 50 keV. The results show that the electron spectrum of 40--100 eV dominates in the low energy range on the sunlit surface in the geomagnetic tail, but a double peak spectrum has been found in the higher energy ranges, including one peak in the range of 300-500 eV and another at 5-6 keV \citep{o1970preliminary}. Subsequently, in situ missions, the Lunar Prospector (LP) and the Acceleration Reconnection Turbulence and Electrodynamics of the Moon's Interaction with the Sun (ARTEMIS) mission, has measured the electron density and temperature, electrostatic potential, and magnetic field. The results show that the electron beam has an energy up to several kiloelectron volts \citep{halekas2005electrons, halekas2009lunarp, halekas2011first, halekas2011new, halekas2014effects}. However, these missions orbited the Moon and made measurements in the lunar orbit at a high altitude, except CPLEE. 

In this study, we focus on the charging and transport characteristics of anorthite particles exposed to beam electrons with an energy of 350 eV. Only incident electrons with an energy of 350 eV were considered here since during the Moon's passage through the terrestrial magnetosphere, where the Moon spends approximately 20\% of its orbit, the increased electron flux density at an energy of 300 to 500 eV has been observed by CPLEE \citep{o1970preliminary}, which means that one of the main plasma components in the magnetosphere is in the equivalent energy range. Moreover, the lunar dust sample is expected to possess the maximum secondary electron yield in the variable energy range of 300--400 eV \citep{anderegg1972secondary, willis1973photoemission, dukes2013secondary}. Here, we directly measure grain diameters and velocities with laser Doppler measurement systems and then calculate the surface potential, the electric field intensity, and the rising heights, aiming at applying the results to the research on dust charging and transport on the lunar surface, especially when the Moon is in the Earth's magnetotail. In addition, the 350-eV electron beam creates a secondary electron sheath if the surface turns out to be positively charged, which could manifest similar properties to a photoelectron sheath on the Moon. Therefore, the results could also be used in analyzing the dust activity over most of the dayside of the Moon, as well as the sloped terrain near the terminator, such as explaining the Surveyor observations of LHG. 

\section{Experimental Technique}

\begin{figure*}
\gridline{\fig{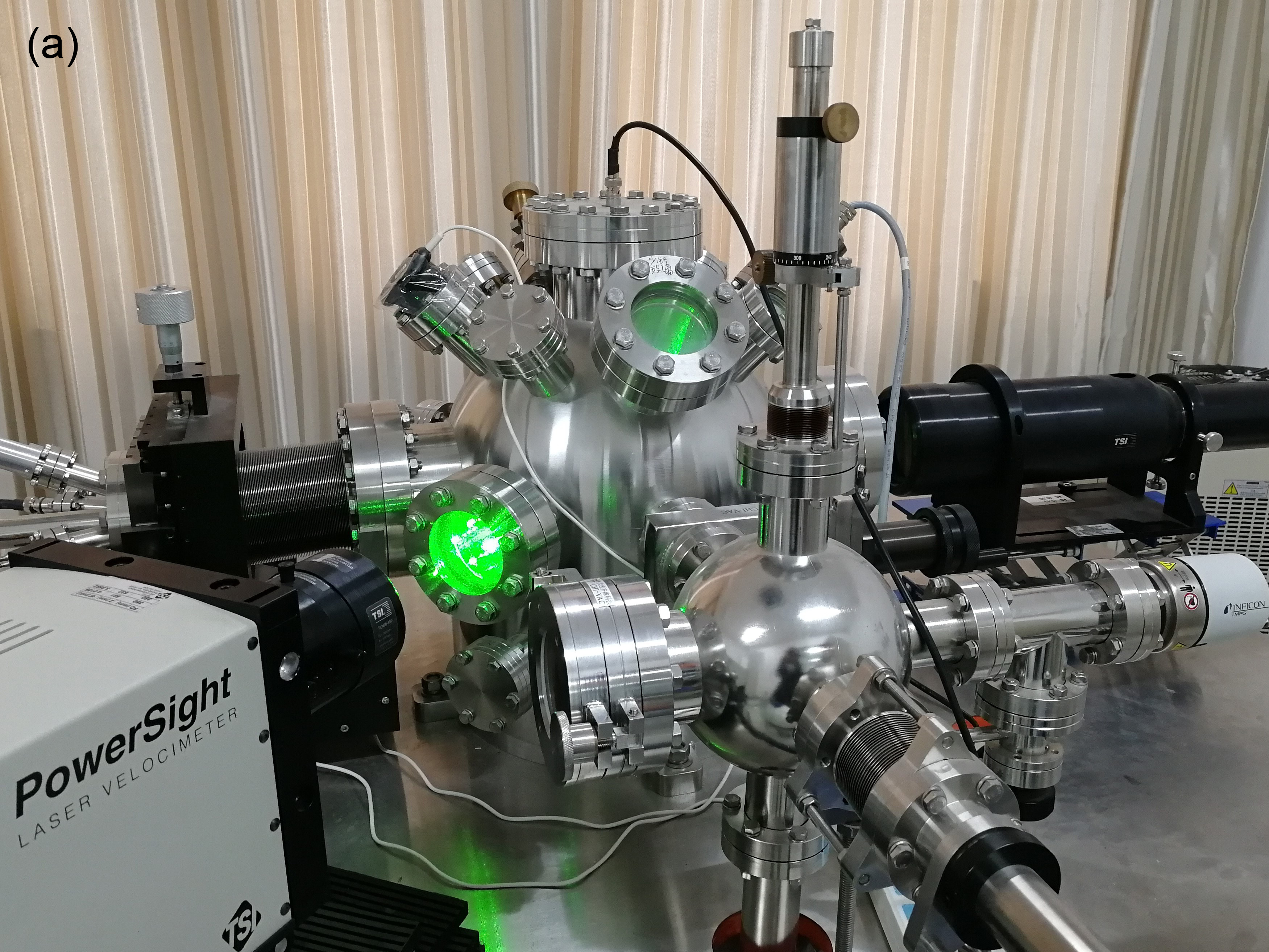}{0.64\textwidth}{}
          \fig{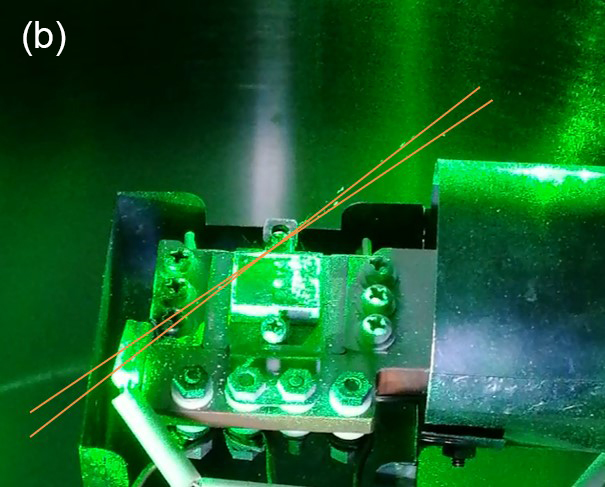}{0.34\textwidth}{}}
\caption{Lunar (Planetary) Dust Environment Simulation Platform: (a) Pictorial view of the experimental apparatus; (b) Image of dust transport. The measured point is at the intersection of two laser beams.
\label{fig:instru}}
\end{figure*}

The Lunar (Planetary) Dust Environment Simulation Platform has been developed since 2013 to measure the diameters and velocities of dust particles lifted in the simulated lunar surface environment, as shown in Figure \ref{fig:instru}. This platform mainly consists of a vacuum system, electron beam system, diameter--velocity (D--V) test system, and data collection system. In the electron beam system, an FS40A-PS Power Supply Unit, which is both a full-option control and a power supply, is employed to control the electron emission. The main operational menu allows the adjustment of characteristic parameters, for example, electron energy in the range of 0--500 eV by steps of 0.1 eV and emission current in the range of 1--500 $\rm{\mu A}$. The electron beam is produced by direct emission from a tungsten cathode ray tube.

In the D--V test system, a phase Doppler particle analyzer/laser Doppler velocimeter (PDPA/LDV) system from TSI (TSI Incorporated, Saint Paul, MN 55126, USA) is used (https://www.tsi.com/products/fluid-mechanics-systems/). Based on the scattering of light by a moving particle, the LDV system can measure the particle velocity. The intersection of two laser beams creates a series of light and dark fringes. The moving particle scatters light when crossing the bright fringe but scatters no light when passing the dark fringe, which results in a fluctuating pattern of scattered light intensity with a frequency ($f_{\rm{D}}$). The velocities of moving particles ($V$) can be calculated by multiplying the frequency by the known fringe spacing ($\delta_{\rm{f}}$):
\begin{equation}
V=\delta_{\rm{f}} \cdot f_{\rm{D}},
\label{eq:v}
\end{equation}
where $\delta_{\rm{f}}$ is determined by the laser wavelength ($\lambda$) and the angle ($2\kappa$) between beams:
\begin{equation}
{\delta _{\rm{f}}}{\rm{ = }}\frac{\lambda }{{2\sin \left( \kappa  \right)}}.
\label{eq:deltaf}
\end{equation}
Note that the frequency of one laser beam was shifted by a Bragg cell by 40 MHz to judge the velocity direction and even to measure velocities near zero. Laser beams are supplied by the PowerSight Laser Velocimeter system with an external beam steering module, which allows the steering of laser beams.

The particle diameter ($D$) can be obtained by the phase Doppler technique, which is an extension of LDV, allowing the size measurements of spherical particles, including liquid sprays as well as some bubbles and solid spheres, to be obtained. The spatial frequency (spacing between the scattered fringes at the light collecting optics) of the scattered fringes can be measured as a phase shift ($\Phi$) between two electrical signals resulting from the scattered light. The diameter is correlated to the phase shift between two laser beams from two corresponding detectors at different positions:
\begin{equation}
D{\rm{ = }}\frac{{L_{\rm{f}}{\delta_{\rm{f}}}\Phi }}{{2\pi \Delta l}}K,
\label{eq:D}
\end{equation}
where $K$ is the optical constant, $ \Delta l$ is the space between two detectors, and $L_{\rm{f}}$ is the lens focal length. In addition, data acquisition and processing were achieved by commercial FLOWSIZER 64 software.

\begin{figure}
\epsscale{0.8}
\plotone{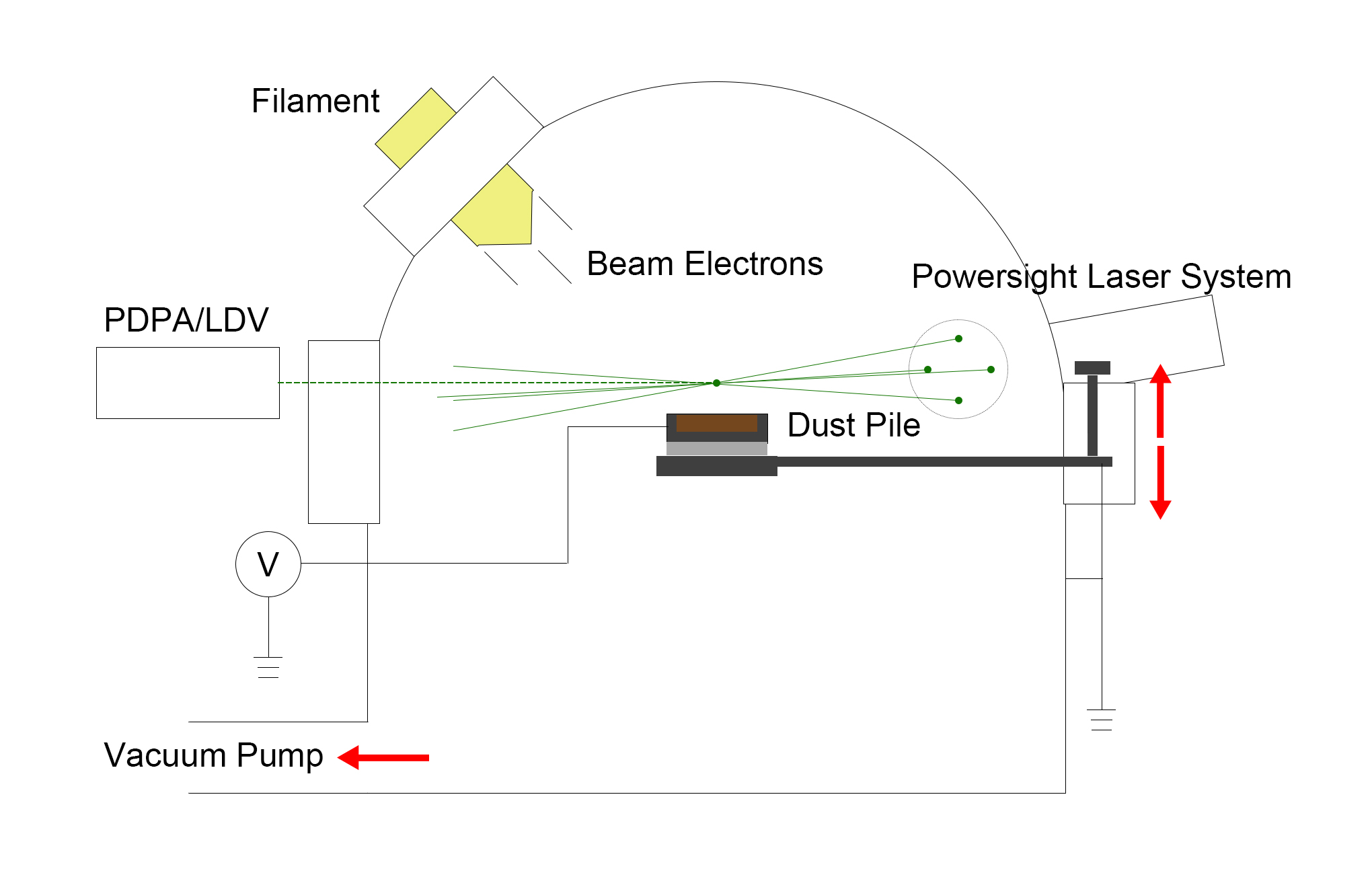}
\caption{Experimental setup. The LDV system allows anorthite particles with velocity in the range of $-$90 to 283 $\rm{m~s^{-1}}$ to be detected; the PDPA system measures the particles with diameter in the range of 0.5 to 90 $\rm{\mu m}$.
\label{fig:setup}}
\end{figure}

As shown in Figure \ref{fig:setup}, the laser source, the PDPA/LDV detector, and the sample stage are deployed in the same horizontal plane, while the filament, the PDPA/LDV detector, and the sample stage are in the same vertical plane. Note that an insulating plate is placed between the sample box and the sample stage. The sample stage and the vacuum chamber are grounded. A voltmeter that directly connects the sample box and the ground is employed to build a real-time measure of the voltage on the sample box. The measured point of dust transport was at the intersection of two laser beams, and the height ($h_0$), namely, the distance between the measured point and the sample surface, can be adjusted by moving the sample stage in the vertical direction. In this experiment, $h_0$ was set to be 6 mm. 

In our study, by using the laser Doppler method, the size and velocity of anorthite particles ejected from a dusty surface can be better characterized than previous studies. The particle size and the vertical velocity were measured simultaneously. The velocity in the range of $-$90 to 283 $\rm{m~s^{-1}}$ with measurement accuracy of 0.2\% can be measured, including the vertical velocity ($V_{\rm{z}}$) and the horizontal velocity ($V_{\rm{x}}$) of the moving particles. The measurable particle size is in the range of 0.5 to 90 $\rm{\mu m}$ with measurement accuracy of 0.5\%. In previous studies, as shown in Table \ref{tab:comp}, the dust transport process was recorded by a high-speed video camera, and dust velocities and diameters were extracted from stacked images \citep{wang2016dust, hood2018laboratory, orger2019experimental, carroll2020laboratory}. However, partial dust migration cannot be detected or tracked by cameras, such as grains whose diameters were lower than several microns (limited by the image resolution), grains that cannot reflect sufficient light consistently, and grains that travel away from the focus range of the camera \citep{orger2019experimental}. For example, the smallest dust particles resolved in Carroll's experiments was approximately 11 $\rm{\mu m}$, and the dust velocities were theoretically calculated, based on the patched charge model and the energy conservation law. Modeling dust migration precisely is difficult because of the randomness of the charging and transport processes of isolated particles. Here, we focus on the group migration behavior of smaller dust grains (0.5--40 $\rm{\mu m}$) based on the statistical particle size and velocity. These particles were similar in size to those putatively observed to have been electrostatically transported on the Moon, e.g., the diameters of dust particle inferred from LHG observed by Surveyor lander TV cameras were approximately 10 $\rm{\mu m}$. Smaller anorthite particles have similar pore sizes and cohesive forces between grains with the lunar regolith. The pore sizes are important for charging mechanisms at the surface, especially for the patched charging model. Second, smaller particles tend to have stronger cohesive forces than larger particles. This means that the results of the experiments are more directly relevant to dust electrostatically transported in the lunar environment. Moreover, a higher electron emission current was used in our experiments to lift more dust grains. 

\begin{deluxetable*}{cccc}
\tablenum{1}
\tablecaption{Comparison of experimental methods and conditions between previous works and this experiment\label{tab:comp}}
\tablewidth{0pt}
\tablehead{
\colhead{Research} & \colhead{experimental methods } & \colhead{Diameter range} & \colhead{Electron energy}
}
\startdata
\citet{wang2016dust}, \citet{carroll2020laboratory} & high-speed video camera & 7-140 $\rm{\mu m}$ & 120 eV \\
\citet{orger2019experimental} & high-speed video camera & 2.4--19$\rm{\mu m}$/pixel & 450 eV \\
This experiment & PDPA/LDV system & 0.5-90 $\rm{\mu m}$ & 350 eV \\
\enddata
\tablecomments{The diameter range represents the range of particle size that can be detected by the corresponding method, rather than the range of particle size measured in the research. }
\end{deluxetable*}

\section{Dust Transport Measurements}

\subsection{Experimental Setup}

Anorthite is one of the major minerals in lunar dust, featuring excellent insulation and a high work function ($\sim$5.58 eV) \citep{li2016indication}. It can acquire charges and even maintain them for a long period. Meanwhile, compared with other lunar surface materials, it may transport easily due to the lower mass density. Therefore, anorthite particles are being treated as a proxy for lunar regolith here. We assume that anorthite particles share similar relevant characteristics to lunar dust, such as shape, secondary emission yields, secondary electron temperatures, etc. In our experiments, anorthite grains were obtained from gabbro (the original rock) in Xigaze, China. The gabbro was crushed preliminarily, and then magnetic substances in the samples were removed by magnetic separation to avoid their influence on the incident electrons and the electrostatic migration of anorthite particles. The collected sample used in our experiments contains $>$90 vol\% anorthite and a small amount of albite. We sieved anorthite particles further to remove particles with sizes greater than 40 $\rm{\mu m}$. Before being measured, the sample was baked at $110{\degr}$C for 24 hours in atmospheric conditions and then deaerated in the vacuum chamber with $10^{-6}$ Pa pressure for 12 hours to remove adsorbed gases.

These experiments were carried out in a hemispherical stainless-steel vacuum chamber, 50 cm in diameter, with a base pressure of approximately $10^{-6}$ Pa. The beam currents ($I_{\rm{e}}$) used in these experiments were fixed at 500 $\rm{\mu A}$ to increase the moving anorthite grain flux. To avoid the effect of a large electric field at the boundary of the beam-illuminated region and the shadow region, as researched by \citet{wang2011dust}, the beam spot size was tuned to 10 mm (the sample box was 8 mm$\times$8 mm). Thus, the beam electron flux ($J_{\rm{e}}$) was approximately 6.4 $\rm{A~{m^{-2}}}$, based on ${J_{\rm{e}}=I_{\rm{e}}/{s}}$, where $s$ is the electron beam spot area. The available incident electron energy was 0--500 eV. However, we found that no significant dust movement was detected under beam electrons below 150 eV, probably because the electric field at the primary electron energy of approximately 150 eV could not push much dust off the surface. From 200 to 500 eV, the response time required for the first dust to leave from the surface gradually decreased but did not significantly change when the incoming electron energy was higher than 300 eV. Here, the measurements under the electron beam with an energy of 350 eV were eventually discussed. The height of the measured point here was 6 mm above the sample surface.

\subsection{Diameter and Velocity}

\begin{figure*}
\gridline{\fig{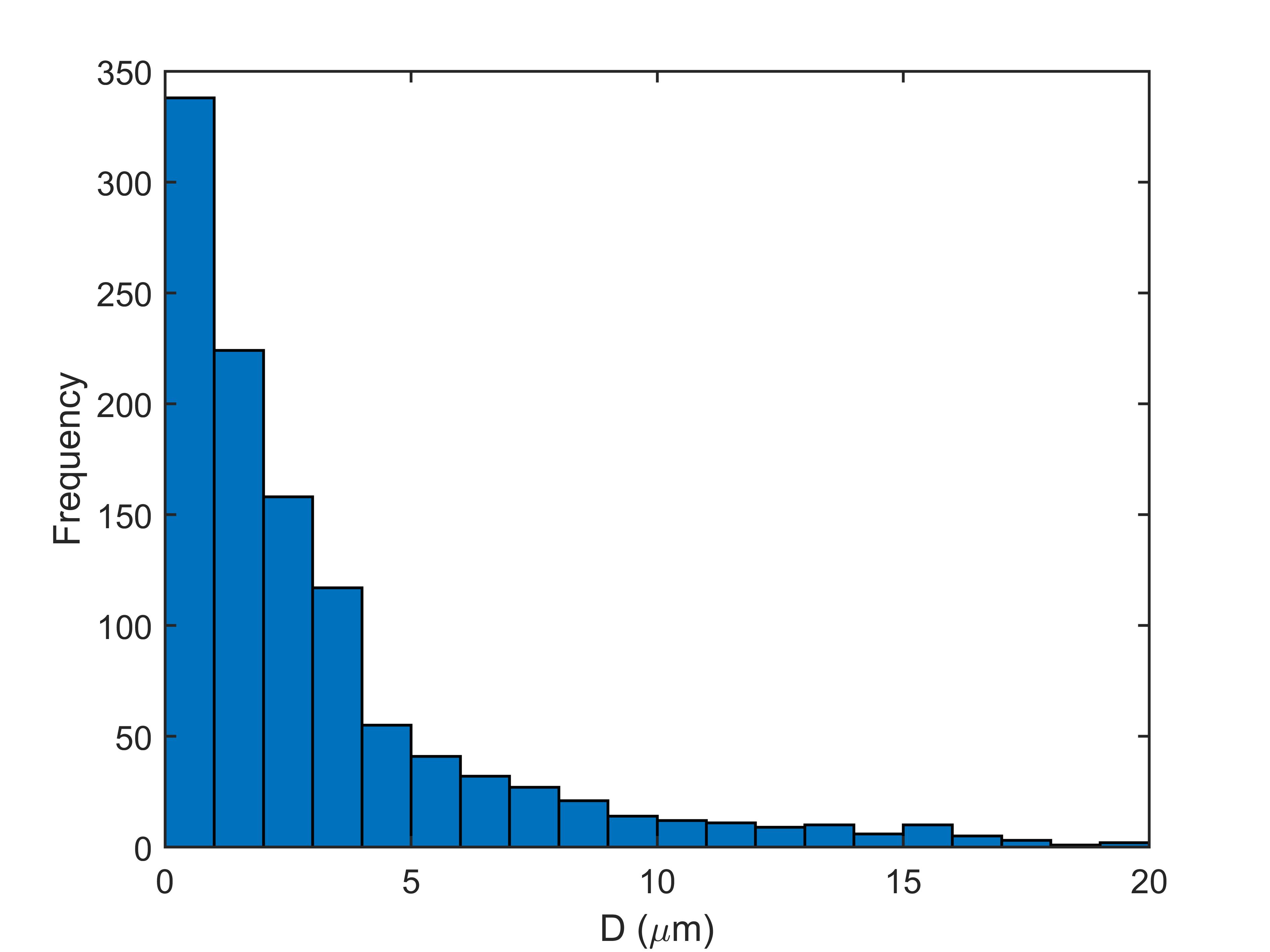}{0.5\textwidth}{(a)}}
\gridline{\fig{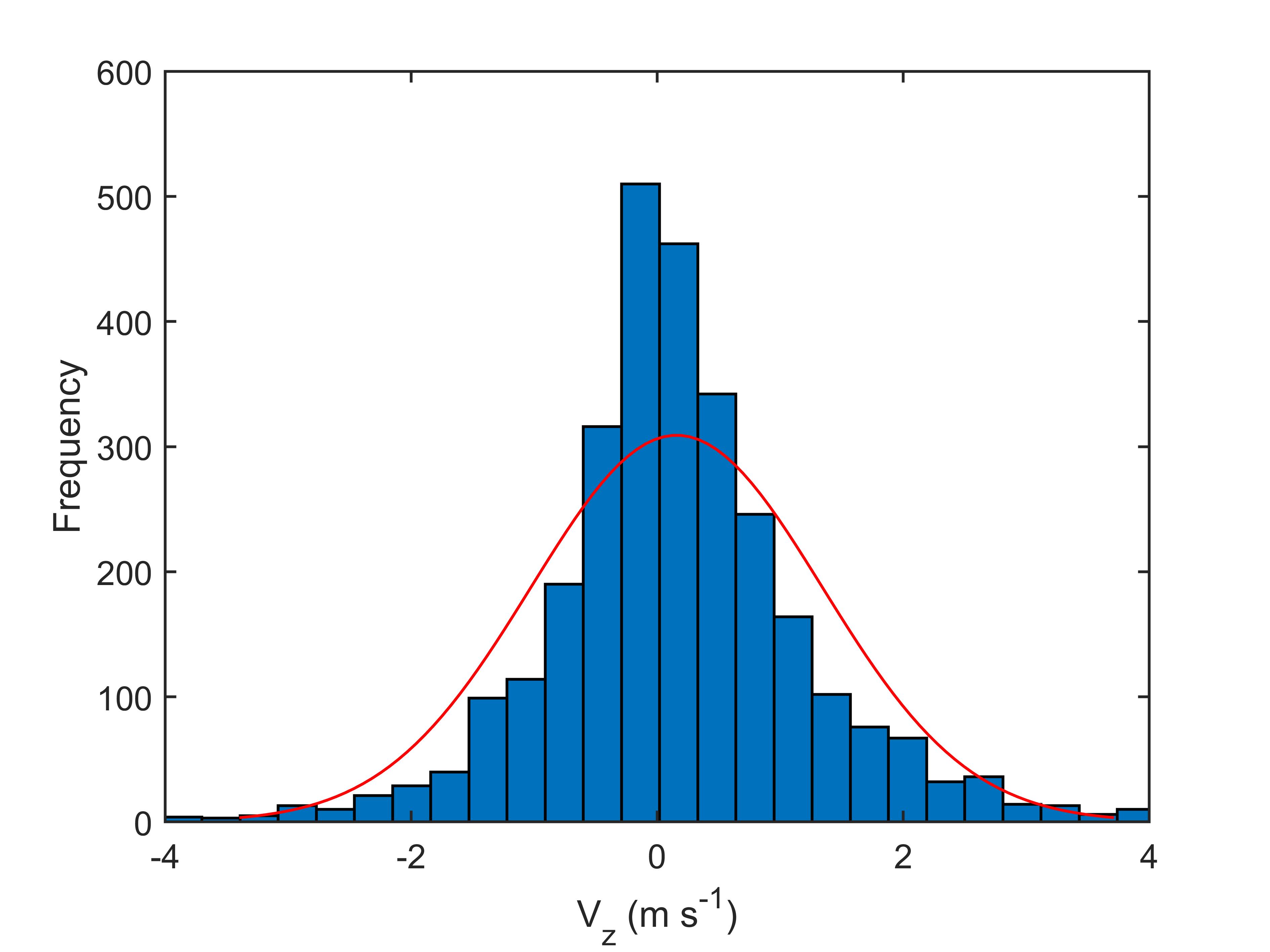}{0.5\textwidth}{(b)}}
\gridline{\fig{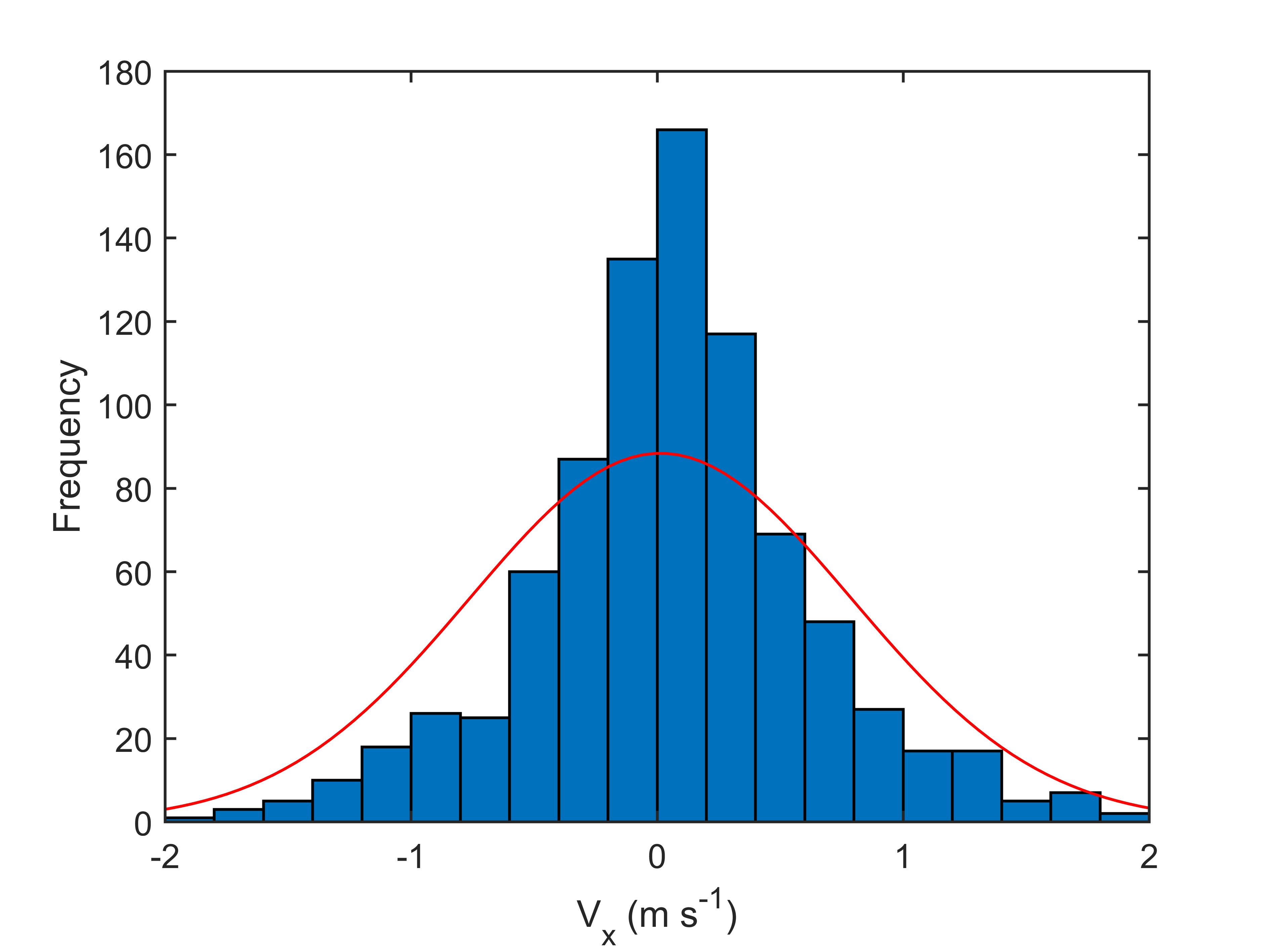}{0.5\textwidth}{(c)}}
\caption{Distributions of the (a) diameter, (b) vertical velocity, and (c) horizontal velocity of lifted anorthite particles under the electron beam with an energy of 350 eV. Gaussian distribution curves of $V_{\rm{z}}$ and $V_{\rm{x}}$ are given by the red lines in (b) and (c).
\label{fig:dvv}}
\end{figure*}

Ten experiments were carried out to measure the diameters and velocities of the moving anorthite particles. Each experiment lasted 5--25 minutes. In all of these experiments, 1125 diameter signals were detected, but there were 20 particles with diameters between 20 and 40 $\rm{\mu m}$ and only 11 particles larger than 40 $\rm{\mu m}$, probably due to particle aggregation. Figure \ref{fig:dvv}(a) shows the diameter distribution. The counts of diameter signals decrease with increasing particle size, and 79.11\% of lifted anorthite particles were smaller than 5 $\rm{\mu m}$.

Both vertical velocities ($V_{\rm{z}}$) and horizontal velocities ($V_{\rm{x}}$) of the moving particles were measured in these experiments. We recorded 2961 vertical velocity signals and 859 horizontal velocity signals, as shown in Figures \ref{fig:dvv}(b) and (c), respectively. $V_{\rm{z}}$ ranges from $-$9.74 to 7.10 $\rm{m~s^{-1}}$, but 91.46\% of them are in the range of $-$2 to 2 $\rm{m~s^{-1}}$. $V_{\rm{x}}$ falls between $-$5.38 and 3.38 $\rm{m~s^{-1}}$, and particles with speeds below 1 $\rm{m~s^{-1}}$ occupy 88.48\%. Gaussian distribution curves of $V_{\rm{z}}$ and $V_{\rm{x}}$ show that there is no significant difference between lifted and falling particles or for particles moving along orthogonal horizontal directions. This indicates that the velocities of anorthite particles are influenced only by gravity and electrostatic forces, and only a miniscule number of particles can reach the chamber ceiling. In other words, the maximum heights of most particles did not exceed 25 cm.

We detected many more signals of the vertical velocity than of the diameter because only one detector is needed to measure velocities, while two detectors are required to obtain particle diameters. Thus, a velocity can be obtained, but the corresponding diameter cannot be obtained if only one detector obtains the signal. On the other hand, the signals of anorthite particle diameters that were below the detection limit of the diameter signal cannot be detected. Furthermore, the number of $V_{\rm{z}}$ signals is greater than that of $V_{\rm{x}}$ signals by a factor of approximately 3.5, since part of the horizontal velocity signals, for example those perpendicular to the $V_{\rm{x}}$ and $V_{\rm{z}}$ directions, cannot be detected. In addition, $V_{\rm{z}}$ falls in a wider range than $V_{\rm{x}}$, though for these particles, gravity needs to be overcome in the vertical direction. The results reveal the dominance of the vertical migration of anorthite particles. The horizontal velocities are mainly ascribed to the effect of the sample box and other nearby grains on the local electric field.

\subsection{Transport Rate of Anorthite Particles}

\begin{figure}
\epsscale{0.65}
\plotone{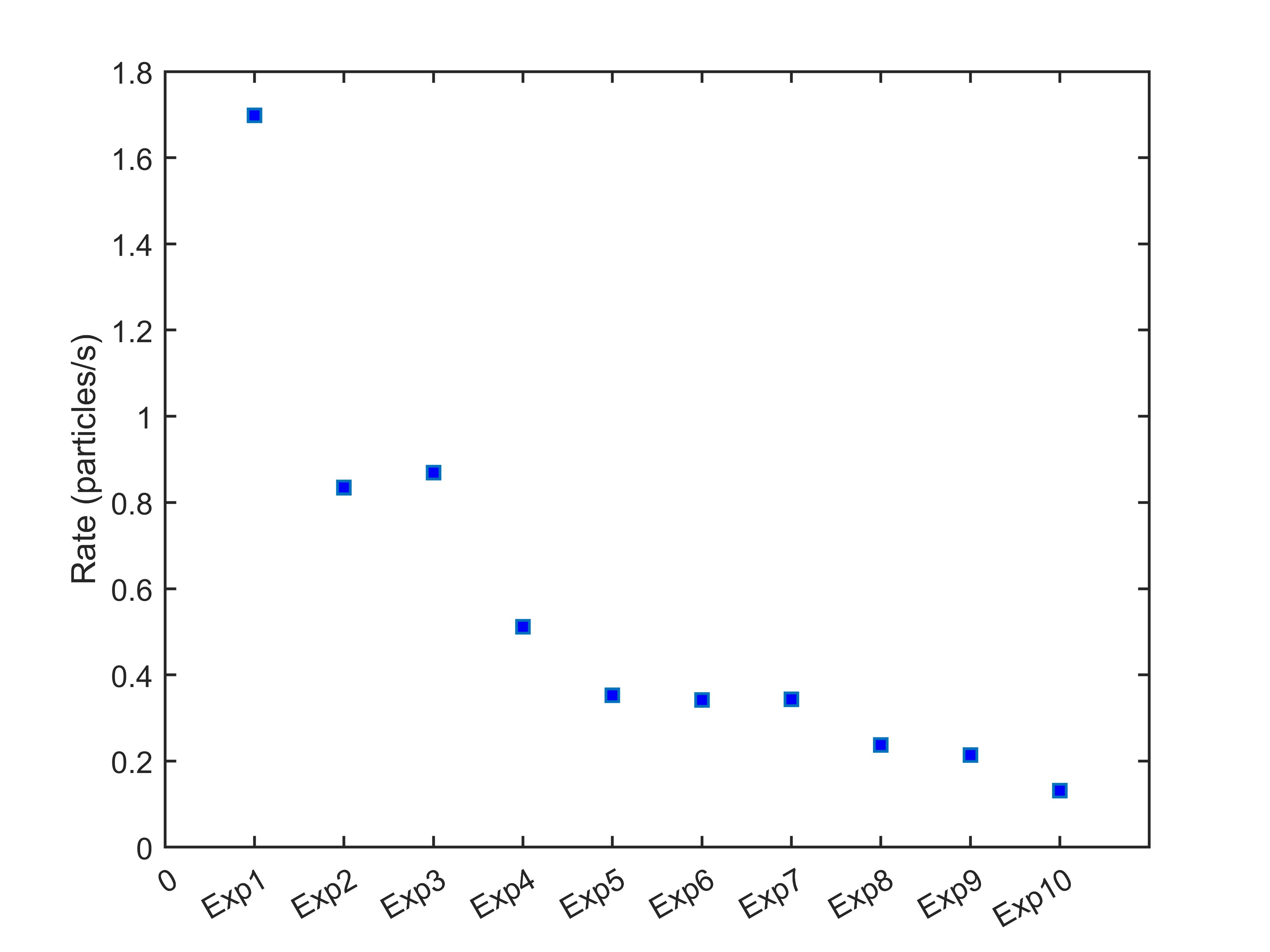}
\caption{Transport Rates of anorthite particles from ten experiments.
\label{fig:rate}}
\end{figure}

To characterize anorthite particle migration, we calculate the average transport rates of lifting particles at the measured point. Counts of cumulative lifted particles are calculated by counting the $V_{\rm{z}}$ signal. Overall, the average rates are mainly limited in the range of 0.1--1.7 particles per second at the measured point. As shown in Figure \ref{fig:rate}, the transport rates decreased with time/depth. Dust activity could be detected with depth, but stronger charging was required to detach the particles due to the increasing packing density and the increasing contact surfaces between the dust grains \citep{orger2019experimental}. This phenomenon is consistent with early experimental results that a lifting rate peaks at the beginning and then slows down \citep{hood2018laboratory}.

Note that this is a type of experiment with sample loss. Anorthite particles in the sample box were reduced since some lifted particles hit the vacuum chamber ceiling and stuck to it, and some lifted particles traveled away from the sample box due to horizontal transport. Anorthite particles in the sample box can be removed completely in a few hours because the thickness of the dust sample is less than 1 mm, so each experiment was set to last from a few minutes to less than thirty minutes.

\begin{figure}
\epsscale{0.65}
\plotone{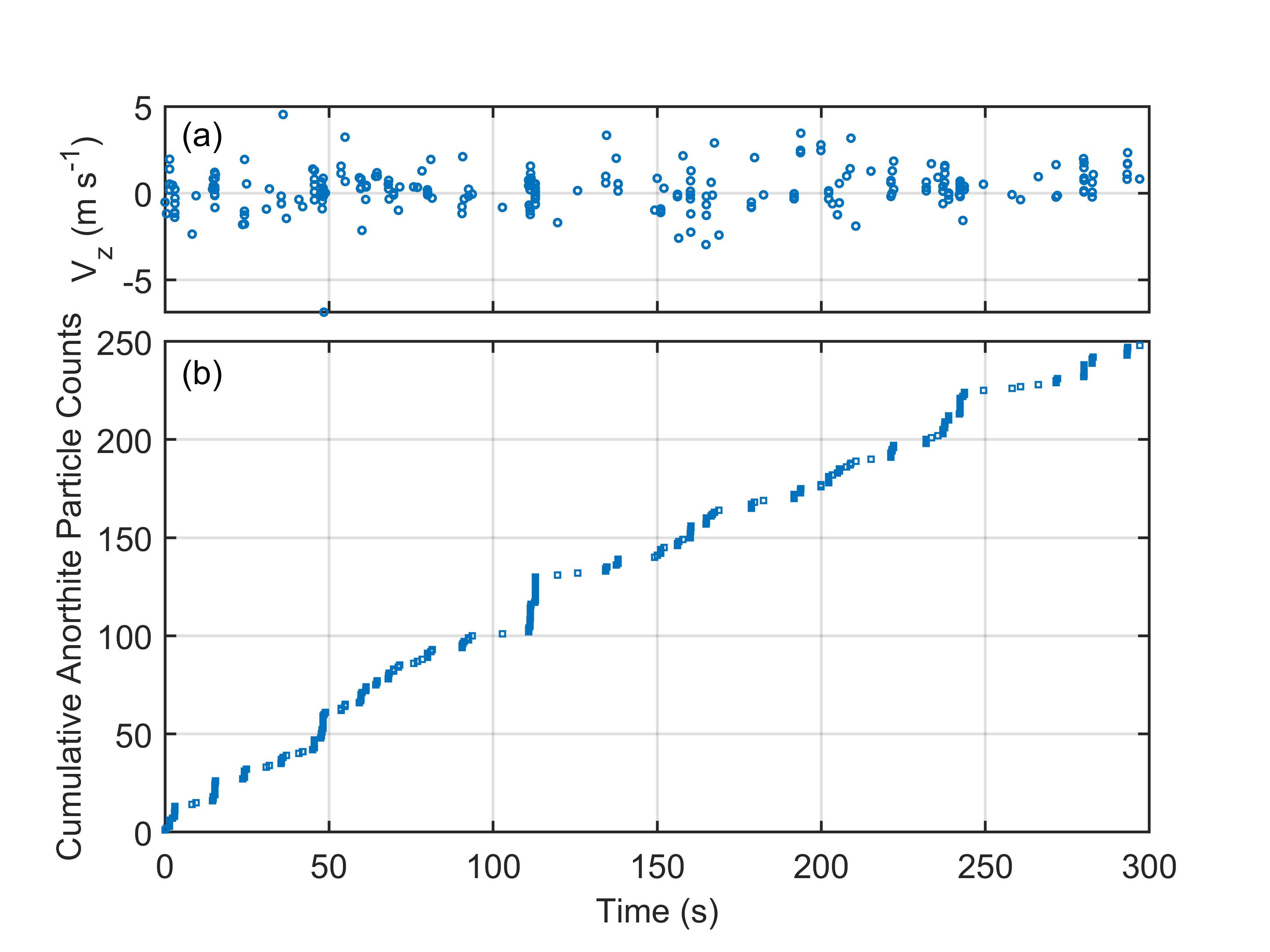}
\caption{(a) The vertical velocity of dust particles as a function of measuring time. (b) The number of cumulative lifted anorthite particles as a function of time based on $V_{\rm{z}}$ signal counts. Note that the time when the first dust particle was liberated was set as the initial time.
\label{fig:d17}}
\end{figure}

The dust migration rates in the first few minutes of each experiment almost maintained constant. As an example, we plot the counts of cumulative lifted particles based on vertical velocities signals. Figure \ref{fig:d17}(a) shows $V_{\rm{z}}$ of lifted dust particles. As mentioned above, the number of upward anorthite particles is approximately the same as that of the downward particles. Particle velocities are mostly in the range of $-$2 and 2 $\rm{m~s^{-1}}$. The number of cumulative lifted particles increases linearly with time, as shown in Figure \ref{fig:d17}(b). The lifting rate is flat during the first 5 minutes, and the average rate is 0.83 particles per second. Interestingly, several or even more than 10 particles were detected almost synchronously. Then after a short time interval ranged from 1 to 12 s, another group of dust particles appeared. This irregular pulse migration is a new phenomenon, implying that the dust particles could be lifted in groups. 

\section{Discussions}

\begin{figure}
\epsscale{0.7}
\plotone{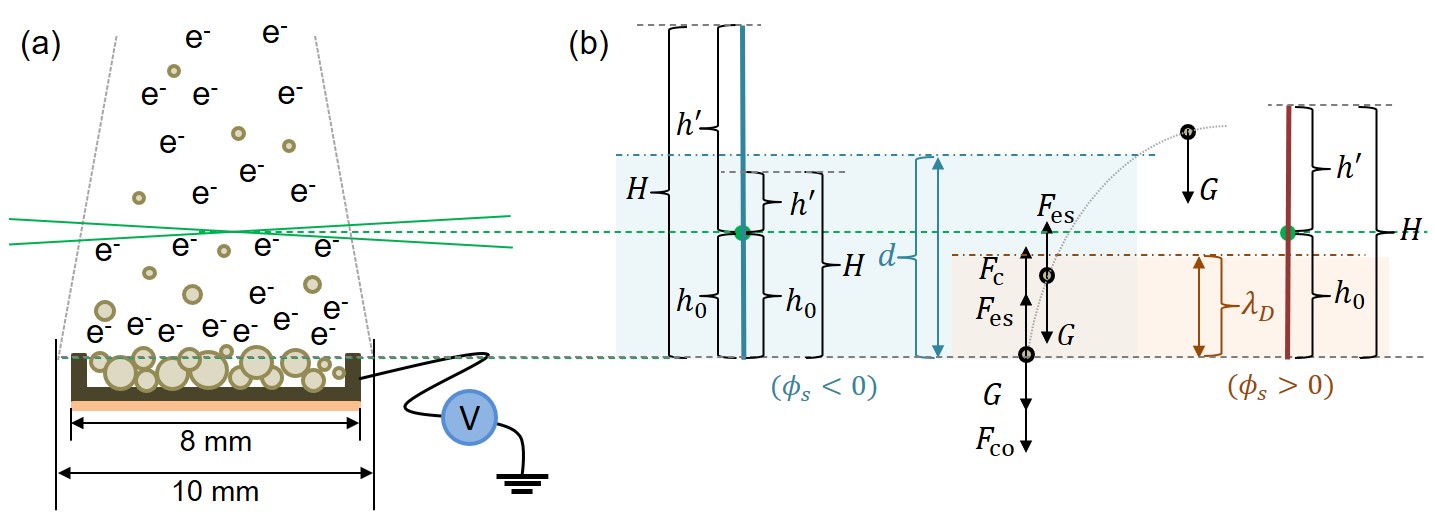}
\caption{(a) Schematic diagram of the experiments; (b) evolution of forces acting on anorthite particles. The green dot in (b) is the measured point corresponding to the intersection of the two laser beams in (a). $h'$ represents the rising height of anorthite particles after reaching the measured point. $\lambda_{\rm{D}}<h_0$ if the sample surface was positively charged; while $d>h_0$ if the surface was negatively charged, and whether the particles could leave the electric field depended on the initial velocities of the particles and their force situation in the field. Note that in the patched charge model, the electric field force was downward if the particle was negatively charged and the sample surface was positively charged. 
\label{fig:force}}
\end{figure}

The experiments were performed below a pressure of $10^{-6}$ Pa. Based on the dynamic fountain model \citep{stubbs2006a} and the recently proposed patched charge model \citep{wang2016dust}, anorthite particles attained charges under electron beam irradiation, and then the electrostatic force ($F_{\rm{es}}$) and the particle-particle repulsive force ($F_{\rm{c}}$) acting on the grains increased. Figure \ref{fig:force} shows the evolution of forces acting on anorthite particles at different heights during dust migration. Once sufficient charges were obtained, that is, $F_{\rm{es}}$ and $F_{\rm{c}}$ became large enough to overcome the gravity ($G$) and the cohesive force ($F_{\rm{co}}$), the particles would be repelled and released by the charged surface with initial velocity ($V_0$). Then, some anorthite particles decelerate in the electrical field and return to the sample surface; and some particles accelerate or decelerate upward through the electrical field, which depends on the magnitude of $F_{\rm{es}}-G$. The particles that can leave the electrical field are decelerated by gravity outside the electrical field and reach the maximum height ($H$). Finally, following a ballistic trajectory, these particles fall back towards the surface. During these processes, the diameters and velocities of lifted anorthite particles were measured at an altitude of 6 mm.

The charging characteristics of anorthite particles and the sample surface determined the electrostatic force and the particle-particle repulsive force acting on anorthite particles as well as the transport characteristics of these particles. Therefore, to further understand the charging and transport characteristics of anorthite particles, we need to first analyze the electric field above the sample surface, such as the equilibrium surface potential and the electric field intensity.

\subsection{Electric Field above the Sample Surface}

In these experiments, anorthite particles were exposed to a 350-eV electron beam without the background plasma. The incident electron beam impacting dust grains is assumed to be well collimated and has initial energy ($E_0$), which can be regarded as a purely monoenergetic beam. To characterize the charging process of the sample surface, we only consider three major charging currents: primary electron current ($I_{\rm{e}}$), secondary electron current ($I_{\rm{se}}$), and backscattered electron current ($I_{\rm{be}}$). Note that the secondary and backscattered electrons from the sample box were neglected.

For an uncharged surface ($\phi_{\rm{0}}=0$), the arriving electron flux ($J_{\rm{e}}$) can be given by
\begin{equation}
{J_{\rm{e}}} = e{n_{\rm{e}}}\sqrt {\frac{{2E_0}}{{{m_{\rm{e}}}}}},
\label{eq:Je}
\end{equation}
where $e$ represents the elementary charge ($1.60\times 10^{-19}$ C). $n_{\rm{e}}$ and $m_{\rm{e}}$ are the electron density and the electron mass ($9.11\times 10^{-31}$ kg), respectively. $n_{\rm{e}}$ was approximately $3.6 \times 10^6$ $\rm{cm^{-3}}$ for the incident electron beam with an energy of 350 eV. For a charged surface with a potential ($\phi_{\rm{s}}$), when beam electrons approach the surface, the electron energy ($E_{\rm{pe}}$) changes and can be determined by $E_{\rm{pe}} = E_0 +e\phi_{\rm{s}}$. Then, the arriving electron flux ($J_{\rm{pe}}$) can be written in the following form \citep{chow1994secondary}:
\begin{equation}
{J_{\rm{pe}}} = e{n_{\rm{pe}}}\sqrt {\frac{{2(E_0 + e{\phi_{\rm{s}}})}}{{{m_{\rm{e}}}}}},
\label{eq:Jpe}
\end{equation}
where $n_{\rm{pe}}$ is the electron density before impacting the charged surface.
Since the incident electron current was fixed, ${J_{\rm{e}}}$ was equal to ${J_{\rm{pe}}}$ when the electron beam was well collimated. Here, we assume that it took a finite time to charge the bulk sample surface and that the surface had reached the equilibrium surface potential ($\phi_{\rm{s}}$) before any grains left from the surface. 

The equilibrium potential could be positive or negative, depending on the yields of secondary electrons and backscattered electrons for the bulk sample surface covered by anorthite particles, $\delta(E_0)$ and $\eta(E_0)$, respectively. For $\delta(E_0)+\eta(E_0)<1$, the sample surface was charged to a negative potential, $\phi_{\rm{s}}<0$. The incoming electrons were repelled and decelerated. Electrons with energy lower than $-e\phi_{\rm{s}}$ could not reach the bulk sample surface. For $\delta(E_0)+\eta(E_0)>1$, the surface was charged to a positive potential, $\phi_{\rm{s}}>0$. The beam electrons were attracted and accelerated. All the electrons can reach the bulk sample surface. However, the excited secondary electrons with energy lower than $e\phi_{\rm{s}}$ could not escape from the range of surface potential. Here, the emitted secondary electrons are assumed to have a Maxwellian velocity distribution, and the backscattered electrons leaving the sample surface are considered to have a similar but somewhat lower energy than they had upon entering, but have an isotropic distribution due to the randomisation of their velocities \citep{whipple1981potentials}. That means the energy of backscattered electrons can be obtained by ${E_{\rm{be}}}=E_0+e\phi_{\rm{s}}$. Therefore, the escaping secondary and backscattered electron flux, namely, $J_{\rm{se}}$ and $J_{\rm{be}}$, can be obtained as follows \citep{stubbs2014dependence}:
\begin{equation}
{J_{\rm{se}}} = -e{n_{\rm{se}}}\sqrt{\frac{{{k_{\rm{B}}}{T_{\rm{se}}}}}{{2\pi{m_{\rm{e}}}}}}\exp(-{\frac {e{\phi_{\rm{s}}}}{{{k_{\rm{B}}}{T_{\rm{se}}}}}})=-\delta(E_0{\rm{ + }}e{\phi_{\rm{s}}}){J_{\rm{pe}}}\exp(-{\frac {e{\phi_{\rm{s}}}}{{{k_{\rm{B}}}{T_{\rm{se}}}}}}), \ \ \ \ (\phi_{\rm{s}}>0)
\label{eq:Jse1}
\end{equation}
\begin{equation}
{J_{\rm{se}}} = -e{n_{\rm{se}}}\sqrt{\frac{{{k_{\rm{B}}}{T_{\rm{se}}}}}{{2\pi{m_{\rm{e}}}}}}=-\delta(E_0{\rm{ + }}e{\phi_{\rm{s}}}){J_{\rm{pe}}}, \ \ \ \ (\phi_{\rm{s}}<0)
\label{eq:Jse2}
\end{equation}
\begin{equation}
{J_{\rm{be}}}=-e{n_{\rm{be}}}\sqrt {\frac{2(E_0{\rm{ + }}e{\phi_{\rm{s}}})}{{{m_{\rm{e}}}}}} =-\eta(E_0{\rm{ + }}e{\phi_{\rm{s}}}){J_{\rm{pe}}},
\label{eq:Jbe}
\end{equation}
where $k_{\rm{B}}$ is the Boltzmann constant ($1.38\times10^{-23}$ $\rm{J~K^{-1}}$). ${n_{\rm{se}}}$, ${T_{\rm{se}}}$ and $\delta(E_0{\rm{ + }}e{\phi_{\rm{s}}})$ are the density, temperature and yield of the secondary electrons escaping from the bulk sample surface covered by anorthite particles, respectively. ${n_{\rm{be}}}$ and $\eta(E_0{\rm{ + }}e{\phi_{\rm{s}}})$ are the density and yield of the emitted backscattered electrons. Note that anorthite is a near-perfect insulator so that no significant current flows through the sample surface or to the sample box. 

(1) Equilibrium Surface Potential

Kirchhoff's law states that all incoming currents are equal to all outgoing currents under equilibrium conditions, which can be used to derive the equilibrium surface potential ($\phi_{\rm{s}}$). That is, the net electron flux satisfies ${J_{\rm{net}}} = 0$. Based on Equations (\ref{eq:Jpe}--\ref{eq:Jbe}), the current balance equation satisfies that:
\begin{equation}
e{n_{\rm{pe}}}\sqrt {\frac{{2(E_0 + e{\phi_{\rm{s}}})}}{{{m_{\rm{e}}}}}} \left[{1 - \delta(E_0 + e{\phi_{\rm{s}}})\exp(-{\frac {e{\phi_{\rm{s}}}}{{{k_{\rm{B}}}{T_{\rm{se}}}}}}) - \eta(E_0 + e{\phi_{\rm{s}}})}\right]= 0, \ \ \ \ (\phi_{\rm{s}}>0)
\label{eq:Jnet1}
\end{equation}
\begin{equation}
e{n_{\rm{pe}}}\sqrt {\frac{{2(E_0 + e{\phi_{\rm{s}}})}}{{{m_{\rm{e}}}}}} \left[ {1 - \delta(E_0 + e{\phi_{\rm{s}}}) - \eta(E_0 + e{\phi_{\rm{s}}})} \right] = 0.
 \ \ \ \ (\phi_{\rm{s}}<0)
\label{eq:Jnet2}
\end{equation}
It means that the current balance can be obtained by a self-consistent physical process among the surface potential ($\phi_{\rm{s}}$) and the yields of secondary electrons ($\delta(E_0 + e{\phi_{\rm{s}}})$) and backscattered electrons ($\eta(E_0 + e{\phi_{\rm{s}}})$). If $\delta(E_0 + e{\phi_{\rm{s}}}) + \eta(E_0 + e{\phi_{\rm{s}}})>1$, the insulating anorthite particles were considered to be positively charged. We assume that ${T_{\rm{se}}}=3~{\rm{eV}}$, $\delta(E_0 + e{\phi_{\rm{s}}})=1.5$ and $\eta(E_0 + e{\phi_{\rm{s}}})=0.1$ \citep{vaverka2016lunar}, thereby the bulk surface potential is $\phi_{\rm{s}}=1.5~{\rm{V}}$. It's worth noting that the different ${T_{\rm{se}}}$ values have a negligible effect on the following results. If $\delta(E_0 + e{\phi_{\rm{s}}}) + \eta(E_0 + e{\phi_{\rm{s}}})<1$ in 0-350 eV energy range, anorthite particles charged negatively. The surface potential could reach $\phi_{\rm{smax}}=-350~{\rm{V}}$. If $\delta(E_0 + e{\phi_{\rm{s}}}) + \eta(E_0 + e{\phi_{\rm{s}}})=1$ at some energy point $E_{\rm{pe}}$ in 0-350 eV energy range, then the surface potential $\phi_{\rm{s}}$ could be obtained by $(E_{\rm{pe}}-E_{\rm{0}})/e$.

(2) Electric Field Intensity

If the surface charged positively, it was shielded by backscattered electrons and secondary electrons. Therefore, the Debye length (${\lambda_{\rm{Di}}}$) related to secondary electrons or backscattered electrons can be expressed as
\begin{equation}
{\lambda _{\rm{Di}}}{\rm{ = }}\sqrt {\frac{{{\varepsilon _0}{k_{\rm{B}}}{T_{\rm{i}}}}}{{{e^2}{n_{\rm{i}}}}}} ,
\label{eq:lambda0}
\end{equation}
where $\varepsilon _0$ is the permittivity of free space ($8.85\times10^{-12}$ $\rm{F~m^{-1}}$). Note that the subscript $i$ represents $be$ when backscattered electrons dominate in the shielding process, or $se$ when secondary electrons dominate. Based on Equation (\ref{eq:Je}--\ref{eq:Jbe}), $n_{\rm{se}}$ and $n_{\rm{be}}$ can be written as follows:
\begin{equation}
n_{\rm{se}}=\delta(E_0+e\phi_{\rm{s}}) n_{\rm{e}}\sqrt{\frac{4\pi{E_0}}{{{{k_{\rm{B}}}{T_{\rm{se}}}}}}},
\label{eq:nse}
\end{equation}
\begin{equation}
n_{\rm{be}}=\eta(E_0+e\phi_{\rm{s}}) n_{\rm{e}}\sqrt {\frac{E_0}{E_0 + e{\phi_{\rm{s}}}}}.
\label{eq:nbe}
\end{equation}
In Equation (\ref{eq:nbe}), the effect of a positive surface potential on the backscattered electron flux was considered negligible, since the energy of backscattered electrons satisfied $E_{\rm{be}}\gg e{\phi_{\rm{s}}}$ when the surface was positively charged, such that they would all escape from a small positive surface potential. With the above assumed emission parameters for secondary and backscattered electrons, we obtain ${\lambda_{\rm{Dse}}=0.9}$ mm and ${\lambda_{\rm{Dbe}}=232.3}$ mm. Since the Debye length is determined by $1/{\lambda_{\rm{D}}^2}=1/{\lambda_{\rm{Dse}}^2}+1/{\lambda_{\rm{Dbe}}^2}=$0.9 mm, this means that the secondary electrons dominate shielding process, while backscattered electrons have a negligible effect on shielding. In this case, the vertical electric field intensity ($E$) was estimated by ${\phi_{\rm{s}}}/{\lambda_{\rm{D}}}=1.7\times$10$^{3}$ $V~m^{-1}$. Note that the Debye sheath is lower than the monitoring position, namely, ${\lambda_{\rm{D}}}<{h_0}$. 

If the surface charged negatively, there is no Debye shielding effect. The electric field could extend to a much larger effective distance $d$, its intensity can be approximately estimated by $E=\phi_{\rm{s}}/d$. Since the size of sample box is 8 mm$\times$8 mm, the effective distance $d$ should be larger than $h_{0}$ of 6 mm, namely $d>h_{0}$.

\subsection{Secondary Electron Emission Characteristics}

Exposed to beam electrons with an energy of 350 eV, the magnitude and polarity of the surface potential of bulk anorthite particles depend on the secondary electron yield. For the individual grains, the maximum secondary electron yield ($\delta_{\rm{M}}$) is approximately 3.2 at the peak energy of $E_{\rm{M}}=350$ eV, based on experimental results on Apollo 11 and 17 dust samples, Minnesota Lunar Simulant (MLS-1), Johnson Space Center (JSC-1) lunar simulant and Lunar Highlands Type (LHT) regolith simulant \citep{horanyi1998electrostatic, nemecek2011lunar}. However, for the bulk dust surface, the maximum yield of 1.4--1.6 has been obtained at approximately 350 eV excitation energy for Apollo 14 and 15 samples \citep{anderegg1972secondary, willis1973photoemission}, while for the Apollo 16 sample, the maximum yield of 1.13 has also been measured at incident energies below 200 eV \citep{dukes2013secondary}. Lunar Prospector (LP) in-situ measurements also show that the secondary electron yield of lunar regolith is lower than the single-particle laboratory results by a factor of 2--3 \citep{halekas2009lunars}. 

The yield reduction is attributable to the scattering and absorption of secondary electrons within microcavities between dust particles, which markedly reduces the escape probability of secondary electrons \citep{anderegg1972secondary, bruining1954physics} and further influences the current balance at the lunar surface and the resulting equilibrium surface potentials \citep{halekas2009lunars}.

\citet{vaverka2016lunar} have discussed the effect of the reduction factors of 2--4 for both the secondary and backscattered electron yields on the lunar surface potentials. The results show that the surface potential oscillates from slightly positive to negative up to $-$50 V for a reduction factor of 2, and higher factors lead to higher negative potentials. In our experiment, the real reduction factor could be related to the stacking structure of anorthite particles. But according to previous experiments, in-situ LP measurements, and the theoretical calculation, the reduction factor of approximately 2--4 for the secondary electron yield of anorthite particles is taken into consideration in our analysis. 

In addition, the backscattered electron yield of a single dust particle increases with the atomic number and the material density, but it always falls between 0.2 and 0.3 for incident electrons with energies of several hundred electron volts. If the reduction factor for the bulk anorthite particles is about 2, taking a typical value of $\delta(E+e\phi_{\rm{s}})=1.5$ and $\eta(E+e\phi_{\rm{s}})=0.1$, a surface equilibrium potential of 1.5 V is obtained according to Equation (\ref{eq:Jnet1}). In our experimental setup, no surface potential measurement is available. We cannot exclude the possibility of a higher reduction factor. For example, if the reduction factor is 4, that is, $\delta(E_0 + e{\phi_{\rm{s}}}) + \eta(E_0 + e{\phi_{\rm{s}}})<1$, the surface could be negatively charged with a maximum negative potential of $-350~{\rm{V}}$. If $\delta(E_0 + e{\phi_{\rm{s}}}) + \eta(E_0 + e{\phi_{\rm{s}}})=1$ at some energy point $E_{\rm{pe}}$, then the surface potential $\phi_{\rm{s}}$ could be obtained by $(E_{\rm{pe}}-E_{\rm{0}})/e$. These results are consistent with the previous calculation that the potential in the magnetosphere could vary from approximately $-$200 V to $+$10 V depending on the SEE yield of the surface \citep{vaverka2016lunar}. Based on Vaverka's calculation, regardless of secondary electron emission, the potentials can reach approximately $-$400 V. Actually, the lunar surface is also observed to charge negatively \citep{halekas2008lunar, halekas2011first, saito2014night}. More in-situ measurements of the near-surface electric field distribution are needed when the Moon is in the terrestrial magnetotail to study the diversity of charging properties at different places. We will also consider to measure the surface potential and the electric field intensity directly in the subsequent experiments. 

\subsection{Dust Transport Dynamics}

(1) Dust Transport Dynamics

Under the irradiation of beam electrons, the electrostatic force and the particle-particle repulsive force are the main driving force for the anorthite particle migration. In the shared charge model \citep{flanagan2006dust}, a uniform charge distribution over the surface of each dust particle is assumed, and the charge of each dust particle with radius ${r_{\rm{d}}}$ is given by
\begin{equation}
q_{\rm{s}}=4\pi{\varepsilon _0}{r_{\rm{d}}}^2E.
\label{eq:qmin}
\end{equation}
If the electric force $q_{\rm{s}}E$ could overcome the gravity and cohesive force, the dust could be lifted. 

In the isolated capacitors model \citep{flanagan2006dust, stubbs2006a}, the dust particle is assumed to be charged as if it were an isolated sphere immersed in a plasma, while the effects of the nearby surface is neglected. As a result, the dust charge is given by
\begin{equation}
q_{\rm{ic}}=4\pi{\varepsilon _0}{r_{\rm{d}}}\phi_{\rm{s}}.
\label{eq:qmax}
\end{equation}
In the patched charge model \citep{wang2016dust}, the emission and re-absorption of photoelectrons and/or secondary electrons at the walls of microcavities formed between neighboring dust particles below the surface could produce large unexpectedly negative charges, the dust charge is given by
\begin{equation}
q_{\rm{p}}=-2\pi{\varepsilon_0}r_{\rm{d}}(\frac{\xi T_{\rm{se}}} {e})=-4\pi{\varepsilon_0}r_{\rm{d}}\phi_{\rm{p}}.
\label{eq:qpartch}
\end{equation}
Here, $\xi$ is an empirical constant between 4 and 10 based on measurements \citep{wang2016dust}. $\phi_{\rm{p}}=\frac{\xi T_{\rm{se}}}{2e}$ could be regarded as an effective charging potential. The charge of dust particles could be enhanced by several orders of magnitudes compared to shared charge model for micron-sized particles. The lift of dust particles are mainly caused by the intense particle-particle repulsive forces $F_{\rm{c}}$ near the surface, which result in a substantial initial emission velocity $V_{0}$ of the dust particle. The dust's initial velocity $V_{0}$ plays an important role in the upward dust transportation. 

In the shared charge model and isolated capacitors model, the electric field force $F_{\rm{es}} = Eq$ is the main driving force. To lift the dust particles, $F_{\rm{es}}$ should be larger than the gravity and cohesive force. The gravity can be obtained by $G={\frac{4}{3}}{\rho}\pi{r_{\rm{d}}}^3g$, where $\rho$ is the density of anorthite particles (2.61 ${\rm{g~cm^{-3}}}$) and $g$ is the gravitational acceleration. Assuming a zero cohesive force and taking $q=q_{\rm{ic}}$, one can calculate the maximum diameter $D_{\rm{max}}$ of anorthite grains which could be lifted. If the surface was positively charged with a potential of 1.5 V, $D_{\rm{max}}$ is 3.3 $\rm{\mu m}$, far below the diameter of the released dust observed in our experiment. Since $q_{\rm{ic}}$ is much larger than  $q_{\rm{s}}$, $D_{\rm{max}}$ could be much smaller than 3.3 $\rm{\mu m}$ according to the shared charge model, which is inconsistent with our diameter measurements. For a negatively charged surface, the surface potential could reach $-$350 V at most. Assuming $q=q_{\rm{ic}}$, $D_{\rm{max}}$ could be larger than 100 $\rm{\mu m}$. Actually, the maximum particle diameter observed in the experiments was also larger than 100 $\rm{\mu m}$. 

During the dust transport process, only the gravity and the electrostatic force worked in the shared charge model and the isolated capacitors model. Based on the energy conservation law, the vertical velocity (${V_{\rm{z}}}$) and the diameter ($D$) satisfy the following relationships:
\begin{equation}
\frac{1}{2}m{V_{\rm{0}}}^2= \frac{1}{2}m{V_{\rm{z}}}^2+mgh_0-Eqh_{\rm{E}},
\label{eq:energ}
\end{equation}
where $h_{\rm{E}}$ is the effective acting range of the electric field. $h_{\rm{E}}={\lambda_{\rm{D}}}$ for the positively charged surface. For the negatively charged surface, we consider that the upper boundary of the electric field is higher than the measured point, according to the discussion in Section 4.1, namely $d>h_0$. That is, $h_{\rm{E}}={h_{\rm{0}}}$ for the negatively charged surface. 

To characterize the relationship between the diameter $D=2{r_{\rm{d}}}$ and the vertical velocity at $h_{0}$, based on Equations (\ref{eq:qmin}-\ref{eq:qmax}) and $m={\frac{4}{3}}{\rho}\pi{r_{\rm{d}}}^3$, Equation (\ref{eq:energ}) can be written as:
\begin{equation}
{V_{\rm{z}}}^2=A{D^{-2}}+C,\ \ \ \ (q=q_{\rm{ic}})
\label{eq:vd1}
\end{equation}
\begin{equation}
{V_{\rm{z}}}^2=B{D^{-1}}+C,\ \ \ \ (q=q_{\rm{s}})
\label{eq:vd2}
\end{equation}
where $A_1= \frac{24{\varepsilon_0}{\phi_{\rm{s}}}^2}{{\rho}}$ for the positively charged surface, $A_2= \frac{24{\varepsilon_0}{h_0}E{\phi_{\rm{s}}}}{{\rho}}$ for the negatively charged surface, $B_1= \frac{12{\varepsilon_0}E{\phi_{\rm{s}}}}{{\rho}}$ for the positively charged surface, $B_2= \frac{12{\varepsilon_0}{h_0}E^2}{{\rho}}$ for the negatively charged surface, and $C=-2gh_0+{V_{\rm{0}}}^2$. Here $V_{\rm{0}}=0$ for the shared charge model and the isolated capacitors model. 

For the patched charge model, the particle-particle repulsive force, electric force and the gravity worked during the dust lifting process. The energy conservation law could be written as:
\begin{equation}
\frac{q_{p}^{2}}{4\pi\varepsilon_{0}D}-\frac{q_{p}^{2}}{4\pi\varepsilon_{0}h_{0}}= \frac{1}{2}m{V_{\rm{z}}}^2+mgh_0-Eqh_{\rm{E}}.
\label{eq:energ2}
\end{equation}
Since $h_{0}\gg D$, the term $\frac{q_{p}^{2}}{4\pi\varepsilon_{0}h_{0}}$ could be neglected. And the term $\frac{q_{p}^{2}}{4\pi\varepsilon_{0}D}=\frac{1}{2}m{V_{\rm{0}}}^2$, which determines the initial velocity $V_{\rm{0}}$ near the surface at $h\gg D$. Then we can obtain $V_{\rm{z}}$ at $h_{0}$:
\begin{equation}
{V_{\rm{z}}}^2=A_{\rm{p}}{D^{-2}}+C_{\rm{p}},\ \ \ \ (q=q_{\rm{p}})
\label{eq:vd3}
\end{equation}
where $A_{\rm{p1}}= \frac{24{\varepsilon_0}{(0.5\phi_{\rm{p}}^{2}-\phi_{\rm{s}}\phi_{\rm{p}})}}{{\rho}}$ for the positively charged surface, $A_{\rm{p2}}= \frac{24{\varepsilon_0}(0.5\phi_{\rm{p}}^{2}+{h_0}E{\phi_{\rm{p}}})}{{\rho}}$ for the negatively charged surface, and $C_{\rm{p}}=-2gh_0$. 

The above equations suggest that the vertical velocity at the monitoring position is a function of the grain diameter. For the patched charge model, the dust's initial velocity $V_{0}$ resulting from the particle-particle repulsive force due to the micro-cavity charging dominates the upward dust transportation if the surface was positively charged ($\phi_{\rm{p}} \gg \phi_{\rm{s}}$), since the electric force due to the sheath electric field was downward. 

\begin{figure}
\epsscale{1.2}
\plotone{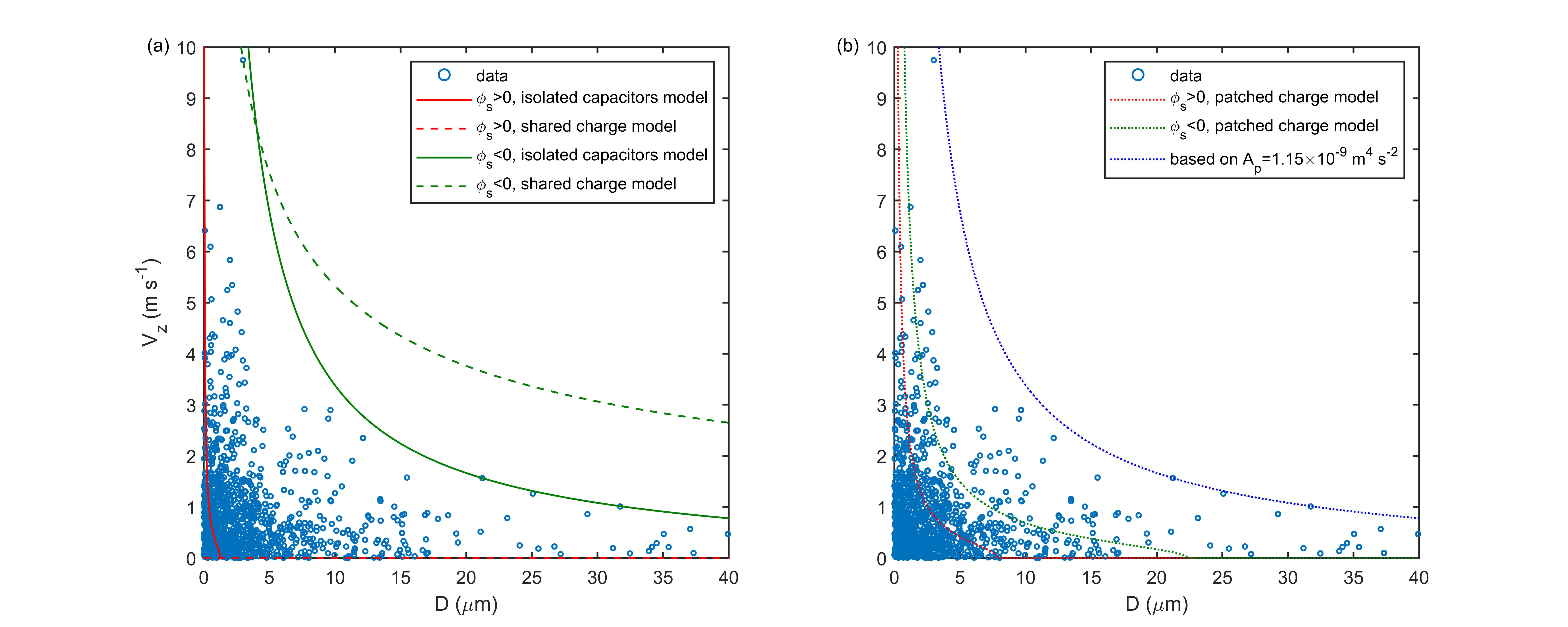}
\caption{Vertical velocity as a function of diameter of the anorthite particle. The red (charged positively) and green (charged negatively) curves are constructed for particles smaller than 40 $\rm{\mu m}$ based on Equations (\ref{eq:vd1}-\ref{eq:vd2}) and (\ref{eq:vd3}). Note that the solid lines show the relationship between ${V_{\rm{z}}}$ and $D$ according to the isolated capacitors model, the dashed lines show the relationship according to the shared charge model, while the dotted lines show the relationship according to the patched charge model ($\xi=10$, $T_{\rm{se}}=3$ eV). In addition, the green solid and dashed lines and the blue dotted line are the upper enveloping curve for particles smaller than 40 $\rm{\mu m}$ based on Equations (\ref{eq:vd1}-\ref{eq:vd2}) and (\ref{eq:vd3}), respectively. 
\label{fig:VD95}}
\end{figure}

In these experiments, the diameters ($D$) and vertical velocities ($V_{\rm{z}}$) of anorthite particles were simultaneously measured at the height of 6 mm. Figure \ref{fig:VD95} shows the distributions of the particle size and their corresponding vertical velocity. As shown in Figure \ref{fig:VD95}, for small particles, especially those with diameters below 5 $\rm{\mu m}$, ${V_{\rm{z}}}$ is distributed in a wider range. Some particles were even accelerated to above 5 $\rm{m~s^{-1}}$. However, for all the particles larger than 15 $\rm{\mu m}$, the speeds were limited within 2 $\rm{m~s^{-1}}$. In addition, anorthite particles with diameters above 40 $\rm{\mu m}$ were probably formed by particle aggregation, whose speeds were always lower than 1 $\rm{m~s^{-1}}$ (not shown here). It is possible that there were still some slow-speed aggregated grains with diameters below 40 $\rm{\mu m}$ released by the surface electric field. The measured diameter for some aggregated grains can be regarded as the equivalent grain size of a new single particle. The experimental results show that the vertical velocities of all the lifted anorthite particles are well constrained by their diameters.

For all the particles with diameters below 40 $\rm{\mu m}$, seven curves in Figure \ref{fig:VD95} are constructed by Equations (\ref{eq:vd1}-\ref{eq:vd2}) and (\ref{eq:vd3}). In the first case, we assume that the surface was positively charged with a surface potential of $\phi_{\rm{s}}=1.5$ V. For the isolated capacitors model, taken $q={q_{\rm{ic}}}$ and ${V_{\rm{0}}}=0$ m/s, the red solid line is given in Figure \ref{fig:VD95}(a), where $A_1$ is 1.83$\times$10$^{-13}$ $\rm{m^4~s^{-2}}$ and $C$ is $-0.12$ $\rm{m^2~s^{-2}}$. While for the shared charge model, the red dashed line in Figure \ref{fig:VD95}(a) is based on ${\phi_{\rm{s}}}>0$, $q={q_{\rm{s}}}$, ${V_{\rm{0}}}=0$ m/s, and then $B_1$ is 1.02$\times$10$^{-10}$ $\rm{m^3~s^{-2}}$. The red dashed curve indicates that based on the shared charge model, all micron-sized anorthite particles cannot migrate. In the isolated capacitors model, anorthite particles with diameters smaller than 3.3 $\rm{\mu m}$ can move upward if we ignore the cohesive force, but only the particles with diameters smaller than 1.3 $\rm{\mu m}$ can reach the measurement point. This is not consistent with our experimental results. 

In the patched charge model \citep{wang2016dust}, anorthite particles acquired a non-zero initial velocity, namely ${V_{\rm{0}}}>0$ m/s, due to the particle-particle repulsive force. Then the particles moved upward and reached the measurement point. Based on the patched charge model, a red dotted line is given in Figure \ref{fig:VD95}(b), using Equation (\ref{eq:vd3}) with $T_{\rm{se}}=3$ eV, $\xi=10$, and $A_{\rm{p1}}=$7.29$\times$10$^{-12}$ $\rm{m^4~s^{-2}}$. Although many data points stay below the curve, there are still a notable fraction of data points with velocities and diameters much larger than the prediction of this curve. To cover all the data points, according to the blue enveloping curve in Figure \ref{fig:VD95}(b), $A_{\rm{p1}}$ is 1.15$\times$10$^{-9}$ $\rm{m^4~s^{-2}}$, then one can get $\xi=113$, much larger than the proposed maximum value of $\xi=10$ observed in the electron beam of 120 eV \citep{wang2016dust}. $\xi=113$ means a charging potential up to $-\xi T_{\rm{se}}/e=-339$ V for these high-speed anorthite particles in the micro-cavity, the absolute value of which is close to $E_{\rm{0}}/e=350$ V. 

It is also possible that the surface was negatively charged. For the negatively charged surface, two upper enveloping curves are constructed by Equations (\ref{eq:vd1}) and (\ref{eq:vd2}), as shown by the green solid and dashed lines in Figure \ref{fig:VD95}(a). Here, ${V_{\rm{0}}}$ is considered to be zero. The green solid line is based on the isolated capacitors model, and $A_2$ is 1.15$\times$10$^{-9}$ $\rm{m^4~s^{-2}}$. Then, we can get $E\phi_{\rm{s}}$, which is 2.36$\times$10$^{6}$ $\rm{V^2~m^{-1}}$. If $\phi_{\rm{s}}=-$350 V, then $E$ and $d$ are 6.75$\times$10$^{3}$ $\rm{V~m^{-1}}$ and 51.80 mm, respectively. Based on the premise that the electric field range is higher than the measured point ($d>h_0$), the maximum surface potential is $-$119 V. Then, $E$ is in the range of 6.75$\times$10$^{3}$--1.98$\times$10$^{4}$ $\rm{V~m^{-1}}$, and the corresponding potential ranges from $-$350 to $-$119 V. While the green dashed line in Figure \ref{fig:VD95}(a) is based on the shared charge model, and $B_2$ is 2.85$\times$10$^{-4}$ $\rm{m^3~s^{-2}}$. Then, $E$ is 1.08$\times$10$^{6}$ $\rm{V~m^{-1}}$, thereby the effective distance of the electric field should be equal to 0.32 mm, which is obviously inconsistent with the premise that $d>h_0$. This means that it is impossible to use the shared charge model to explain the migration of these high-speed anorthite particles. 

According to the patched charge model \citep{wang2016dust}, the green dotted line, as shown in Figure \ref{fig:VD95}(b), is obtained with $T_{\rm{se}}=3$ eV, $\xi=10$ and $E=6.75\times10^{3}$ $\rm{V~m^{-1}}$, namely $A_{\rm{p2}}=$5.86$\times$10$^{-11}$ $\rm{m^4~s^{-2}}$. Some of the data sites outside the line, which means a higher $\xi$ or $E$ is needed. $Eh_0$ in the expression for $A_{\rm{p2}}$ represents the potential difference between the measured point and the sample surface. According to the blue enveloping curve as shown in Figure \ref{fig:VD95}(b), $A_{\rm{p2}}$ is 1.15$\times$10$^{-9}$ $\rm{m^4~s^{-2}}$. Here, we consider the possible potential difference to be $-$350, $-$250, $-$150, $-$50 V or 0 V, then the corresponding $\xi$ are 26, 34, 50, 84 or 112, respectively. The $\xi$ values extracted from both the positive and negative surface charging conditions are larger than the proposed maximum value of $\xi=10$ observed in the previous experiment with an electron beam of 120 eV \citep{wang2016dust}. 

There may exist possible correlation between various $\xi$ values and incident electron beam energies and fluxes. In the micro-cavity, there are backscattered electrons besides the secondary electrons. If the source particles corresponding to the secondary electrons emission is positively charged, then the flux of secondary electrons could be significantly suppressed according to Equation (\ref{eq:Jse1}). Due to the high emission energy of backscattered electrons, the flux of backscattered electrons in the cavity is slightly decreased. In this case, the backscattered electrons could be the main charging current in certain micro-cavity in our experiment because of substantial backscattered electron yield of 0.2$-$0.3 at 350 eV. For a certain anorthite particle, it is possible that $\delta(E_0 + e{\phi_{\rm{s}}}) + \eta(E_0 + e{\phi_{\rm{s}}})<1$ due to fluctuations in its chemical composition or shape. As a result, a larger charging potential up to $-$350 V is possible. This senario is consistent with a large $\xi$ value of 26--113 (corresponding $-\xi T_{\rm{se}}/e=-$339--$-$78 V) extracted from our experiments. By assuming the maximum charging potential for a levitating anorthite particle, it is not surprising that the isolated capacitors model could describe the data in the case of a negatively charged surface. In this case, $q_{\rm{ic}}$ is the upper limit of $q_{\rm{p}}$. But the isolated capacitors model could not work if the surface was positively charged since in this case $\phi_{\rm{s}}$ is too small compared to the negative surface potential. It will be very interesting to measure the surface potential in the later experiment to verify different model scenarios. 

In a word, the isolated capacitors model and the patched charge model can effectively explain the current experimental data if the surface was negatively charged. In addition, the data below the blue enveloping curve denotes that $q<{q_{\rm{ic}}}$ or $q<{q_{\rm{p}}}$, that is, before being launched, a lot of particles did not obtain the saturated charge, which constrains the magnitude of the cohesive force between anorthite particles. 

When $F_{\rm{c}}$ is the main driving force, the maximum size of lifted anorthite particles on the lunar surface may increase due to the reduced lunar gravity. When $F_{\rm{es}}$ is the main driving force, the maximum size of lifted anorthite particles on the lunar surface strongly depends on the electric field characteristics and the cohesive force between particles. The electrostatic field on the lunar surface could be lower than that in this experiment. This is caused by the lower incident electron current on the Moon, which means that the Debye length is significantly higher than that in our experiments. Generally, the Debye length on the Moon varies from dozens of centimeters to hundreds of meters. Based on Equations (\ref{eq:lambda0}) and (\ref{eq:nse}), for example, when the incident electron current on the Moon is reduced by 10000 times, the Debye length is increased by 100 times (0.17 m). However, it would become shorter for the surface with a higher secondary electron emission yield, especially in the space environment with a high electron flux and/or high electron energy, such as during the Moon's passage through the geomagnetic tail and during the solar burst period. 

(2) Migration Height

\begin{figure}
\epsscale{0.65}
\plotone{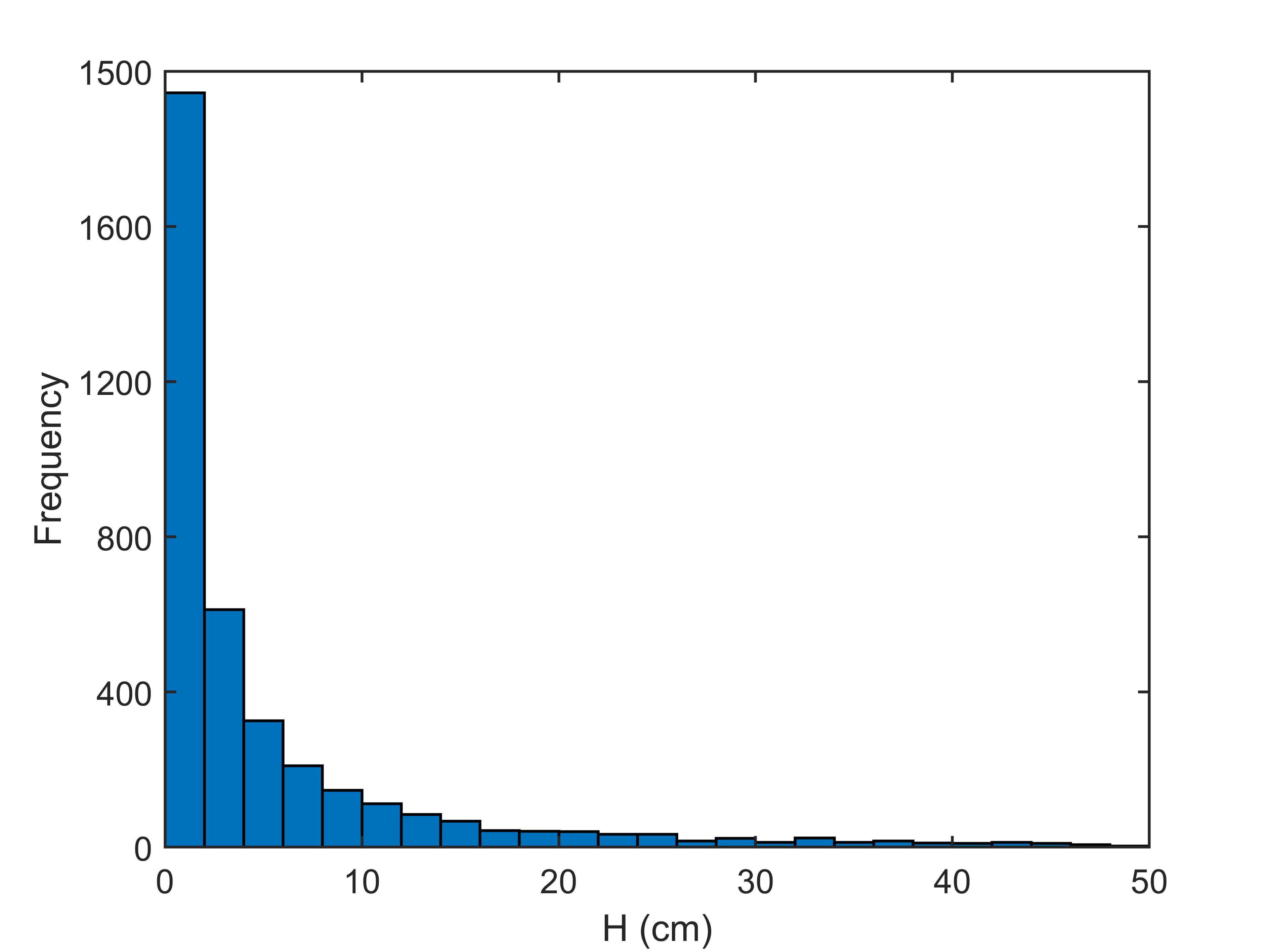}
\caption{Height distribution of lifted anorthite particles based on Eqation \ref{eq:H1}, where $V_{\rm{z}}$ was measured at the height of $h_0$ in our experiments. Note that $h_0$ was higher than $\lambda_{\rm{D}}$.
\label{fig:H}}
\end{figure}

When $\phi_{\rm{s}}>0$, $\lambda_{\rm{D}}<h_0$, based on the energy conservation law, the maximum rising height $H$ of anorthite particles can be written as:
\begin{equation}
H = {\frac{{{{V_{\rm{z}}}^2}}}{{2g}} + {h_0}}.
\label{eq:H1}
\end{equation}
When $\phi_{\rm{s}}<0$, $d>h_0$, for the particles that decelerated ($Eq<mg$) and cannot leave the electric field ($H<d$), the maximum rising height is
\begin{equation}
H = {\frac{{{{V_{\rm{z}}}^2}}}{{2g}}\cdot {\frac{mg}{mg-Eq}}+ {h_0}}.
\label{eq:H2}
\end{equation}
And for the particles that can leave the electric field range ($H>d$), the maximum rising height is
\begin{equation}
H = {\frac{{{{V_{\rm{z}}}^2}}}{{2g}} + {h_0}}+\frac{Eq}{mg}(d-h_0).
\label{eq:H3}
\end{equation}
The maximum rising height $H$ increases with the observed vertical velocity at the height of $h_0$. Figure \ref{fig:H}, for example, presents the distribution of $H$ derived from Equation (\ref{eq:H1}). The dust flux rapidly decreases with height. Nearly 70\% of the moving particles cannot reach a height of 5 cm. Particles with a migration height below 2 cm take up almost half of all lifted anorthite particles. It is worth noting that if the surface was negatively charged, the migration height will be even larger, according to Equations (\ref{eq:H1}-\ref{eq:H3}), than that when the surface was positively charged. Nonetheless, the results are supported by our measurements that numerous particles visible to the naked eye can only migrate to the heights of a few centimeters above the sample surface. In addition, 131 particles ($\sim$4.4\%) rose to heights above 25 cm, which impacted the chamber's ceiling. Half of them probably adhered to the ceiling or were scattered to other places, since only 70 particles ($\sim$2.4\%) that had reached heights above 25 cm, based on Equation (\ref{eq:H1}), returned the sample surface.

On the lunar surface, dust grains can reach higher positions because of the lower gravitational acceleration. Based on Equation (\ref{eq:H1}), with one-sixth of Earth's gravitational acceleration ($g_{\rm{L}}=1/6g$) and identical electric field characteristics, the equivalent lifting height on the lunar surface is approximately six times that on Earth if dust grains satisfy $F_{\rm{es}}+F_{\rm{c}}>mg_{\rm{L}}+F_{\rm{co}}$. That is, only approximately 30\% of lifted particles can reach a height higher than 30 cm, while approximately 70\% of lifted particles migrate at a height below 30 cm. This result is consistent with the recent conclusion of a dust deposition upper line on rocks of approximately 28 cm \citep{yan2019weak}, and also matches well with the Surveyor LHG observations.

(3) Pulse Migration

The pulse migration of anorthite particles was found in these experiments. Several anorthite particles were detected almost simultaneously. The velocities of different anorthite particles are different from each other. The measurement point $h_{0}$ is 6 mm above the surface. Taking two typical emission velocities of 0.1 m/s and 0.5 m/s as an example, this implies that the emission time difference of the two anorthite particles are within 50 ms. It is not clear whether these anorthite particles were emitted at the same place, but it is possible in terms of the charging time of each dust particle. With a steady incident current ($I_{\rm{net}}$), the charging time ($\tau$) for a spherical particle can be given by $\tau I_{\rm{net}} = q$ \citep{lai2011fundamentals}, which can be obtained by $q=4\pi{\varepsilon _0}{r_{\rm{d}}}\phi_{\rm{g}}$ and $I_{\rm{net}}=J_{\rm{net}} \pi {D^2}/4$, where $\phi_{\rm{g}}$ is the grain's effective charging potential. $\phi_{\rm{g}}$ could be $\phi_{p}$ or $\phi_{s}$, depending on different models. Thus, the charging time for anorthite particles can be expressed by
\begin{equation}
\tau {\rm{ = }}\frac{{8{\varepsilon _0}{\phi_{\rm{g}}}}}{{{J_{\rm{net}}}D}}.
\label{eq:tau}
\end{equation}
The above equation shows $\tau \propto {D^{ - 1}}$ under the same incident flux condition. More time is needed to charge smaller anorthite particles to the equilibrium potential. Certainly, these particles may have been liberated before charging to saturation charges, that is, it takes less time than $\tau$. Based on Equation (\ref{eq:tau}), for example, in a vacuum environment of flux density 5 $A~{m^{-2}}$, a spherical surface of $R=1$ $\rm{\mu m}$ could be charged from 0 to 300 V in approximately 2 ms. Therefore, it is possible that anorthite particles in the same place were emitted one by one in a short time duration. 
In addition, the interval between two pulse migrations ranged from 1 to 12 seconds in our experiments, which is distinctly longer than the emission time duration of the anorthite particles within one pulse migration.

The top surface may contain many dust aggregates, it is possible that the emission of one anorthite particle in the aggregates will decrease the cohesive forces of its neighboring particles, since the number of contacting particles for these neighboring particles could decrease due to the emission of this particle. Therefore, the emission of one anorthite particle could lead to a chain reaction, causing the emission of other particles in the aggregate in a short time duration, that is, pulse migration. A further detailed study of the micro-structure characteristics of the emission position of these pulse migrations may shed light on the emission mechanisms of dust particles. For example, one can check whether there are micro-cavities associated with the dust emission position as proposed by the patched charge model \citep{wang2016dust}. 

\section{Conclusions}
To understand the interaction of incident electrons and dusty surface as well as the dust dynamics on the lunar surface, we measured the diameters and velocities of moving anorthite particles under electron beam irradiation with an energy of 350 eV for the first time. The measurements confirmed the vertical and horizontal migration of anorthite particles. Approximately 79\% of anorthite particles liberated by the electric field were smaller than 5 $\rm{\mu m}$. At the measurement point of 6 mm above the dusty surface, the speeds of approximately 90\% particles were lower than 2 m~s$^{-1}$ in the vertical direction and lower than 1 m~s$^{-1}$ in the horizontal direction. The average transport rates were principally in the range of 0.1--1.7 particles per second. Meanwhile, the pulse migration of multiple anorthite particles within very short-time duration (usually microseconds to tens of microseconds) was also observed. Further analysis about the surface micro-structures characteristics associated with this pulse migration may shed light on the dust emission mechanisms. 

In our experiments, some anorthite particles were observed to have a large vertical velocity up to 9.74 m~s$^{-1}$ at the measurement point. The upper boundary of the vertical velocities $V_{\rm{z}}$ of these high-speed anorthite particles are well constrained by its diameter $D$, that is, $V_{\rm{z}}^2$ linearly depends on $D^{-2}$. To understand this relationship, we have discussed the surface potential, electric field intensity, and dust transport dynamics based on various models. The calculations reveal that the polarity of the bulk surface potential mainly depends on the secondary electron emission yield of the bulk sample surface, which is reduced by 2--3 times relative to that of an isolated particle. If the surface was charged positively, a typical surface potential was estimated to be 1.5 V. If the surface was charged negatively, the surface potential was not less than $-$350 V. The shared charge model could not explain the observed velocity-diameter data in our experiments no matter the surface was positively or negatively charged. The isolated capacitors model also largely underestimates the $V_{\rm{z}}$ and $D$ of the emitted anorthite particles if the surface was positively charged. But it could explain the emission velocity of these high-speed anorthite particles if the surface potential $\phi_{\rm{s}}$ is estimated to be between $-$350 and $-$119 V. Besides the above-mentioned two surface electric field acceleration model, the patched charge model, in which additional particle-particle repulsive force in the dust micro-cavity are considered, could explain the data no matter the surface was positively charged or negatively charged. However, a much larger effective charging potential ($-\xi T_{\rm{se}}/e$) of $-$339--$-$78 V for the dust charging in the micro-cavities should be utilized to reproduce the $V_{\rm{z}}$ and diameters data of the high-speed anorthite particles. Secondary electrons charging mechanism in the micro-cavities could be very hard to produce so large charging potential. For an incident electron energy of 350 eV, our experimental data may suggest an alternative charging mechanism from backscattered electrons which could produce much larger surface charges in the micro-cavity and stronger particle-particle repulsive force. In a word, the velocity-diameter data observed in our experiments provide strong constraints on various mechanisms related to dust charging and transportation. It would be very helpful to use these data to understand the dust charging and electrostatic transport mechanisms in airless celestial bodies such as the Moon and asteroids.

\acknowledgments
We acknowledge the National Natural Science Foundation of China (Nos. 41903058, 41931077, 11761161001) and the Science and Technology Development Fund (FDCT) of Macau (Nos. 007/2018/AFJ, 008/2017/AFJ, 0042/2018/A2). This work was also supported by the Strategic Priority Program of the Chinese Academy of Sciences (No. XDB41020300), the Youth Innovation Promotion Association of the Chinese Academy of Sciences, the Key Research Program of the Chinese Academy of Sciences (No. XDPB11), the Beijing Municipal Science and Technology Commission (No. Z181100002918003) and the Scientific Research Program of Guizhou Institute of Technology (No. XJGC20181290).

\bibliography{Experiments_on_anorthite-manuscript}{}
\bibliographystyle{aasjournal}
%\nolinenumbers
\end{document}